\documentclass[notitlepage,aps,prd,reprint,superscriptaddress,showpacs]{revtex4-2}

\usepackage{amsmath, amssymb, graphicx, setspace}
\DeclareMathOperator{\sign}{sign}

\usepackage{commath}
\usepackage{mathtools,bm, tabularx, booktabs}
\usepackage{float}
\usepackage{xcolor}
\usepackage[normalem]{ulem}
\usepackage{subfigure}
\usepackage{siunitx}
\usepackage[colorlinks=true]{hyperref}

\usepackage[nolist,printonlyused,nohyperlinks]{acronym}
\newcommand{\mathsym}[1]{{}}
\newcommand{\unicode}[1]{{}}

\newcommand{\rtHz}{\ensuremath{\sqrt{\text{Hz}}}}

\usepackage[capitalise,nameinlink,english]{cleveref}
\usepackage{adjustbox}
\usepackage{physics}

\begin{document}

\title{The influence of Laser Relative Intensity Noise in the Laser 
Interferometer Space Antenna (LISA)}

\def\addressb{Max Planck Institute for Gravitational Physics (Albert-Einstein-Institut), 
30167 Hannover, Germany}
\def\addressluh{Leibniz Universit\"at Hannover, 30167 Hannover, Germany}
\def\addressa{European Space Astronomy Centre, European Space Agency, Villanueva de la
    Ca\~{n}ada, 28692 Madrid, Spain}
\def\addressc{APC, Universit\'e de Paris, CNRS, Astroparticule et Cosmologie, F-75006 
    Paris, France}

\def\addresscc{SYRTE, Observatoire de Paris-PSL, CNRS, Sorbonne 
Universit\'e, 
LNE, Paris, France}

\def\addressm{The UK Astronomy Technology Centre, Royal Observatory, 
Edinburgh, Blackford 
    Hill, Edinburgh, EH9 3HJ, UK}

\def\addressr{University of Glasgow, Glasgow G12 8QQ, United Kingdom}

\author{L~Wissel}\affiliation{\addressb}\affiliation{\addressluh}
\author{O~Hartwig}\affiliation{\addresscc}
\author{J\,B~Bayle}\affiliation{\addressr}
\author{M~Staab}\affiliation{\addressb}\affiliation{\addressluh}
\author{E\,D~Fitzsimons}\affiliation{\addressm}
\author{M~Hewitson}\affiliation{\addressb}\affiliation{\addressluh}
\author{G~Heinzel}\affiliation{\addressb}\affiliation{\addressluh}

\date{\today}

\begin{abstract}
The \ac{LISA} is an upcoming ESA mission that will detect gravitational waves in space by interferometrically measuring the separation between free-falling test masses at picometer precision. To reach the desired performance, \ac{LISA} will employ the noise reduction technique \ac{TDI}, in which multiple raw interferometric readouts are time shifted and combined into the final scientific observables. Evaluating the performance in terms of these \ac{TDI} variables requires careful tracking of how different noise sources propagate through \ac{TDI}, as noise correlations might affect the performance in unexpected ways.
One example of such potentially correlated noise is the \ac{RIN} of the six lasers aboard the three LISA satellites, which will couple into the interferometric phase measurements. 
In 
this article, we calculate the expected \ac{RIN} levels based on the current mission 
architecture and the envisaged mitigation strategies. We find that strict requirements on 
the technical design reduce the effect from approximately 
$\SI{8.7}{pm/\rtHz}$ per inter-\ac{SC} interferometer to that of a much 
lower sub--$\SI{1}{pm/\rtHz}$ noise, 
with typical characteristics of an uncorrelated readout noise after 
\ac{TDI}. Our investigations underline the importance of sufficient balanced 
detection of the interferometric measurements.
\end{abstract}


\maketitle

\section{Introduction}
The \acf{LISA} is a future space mission that will detect  
gravitational waves in the mHz range \cite{LISAL3, LISA_req_doc2018}. 
It consists of a constellation of three identical \acf{SC}, each of which 
follows a heliocentric orbit at similar 
distance to the Sun as the Earth, such that the whole constellation forms an almost equilateral triangle either leading or trailing our planet with an 
angular 
separation of $\SIrange{10}{30}{\degree}$. Each spacecraft hosts two free-falling 
\acp{TM}, which are shielded inside the 
\acp{SC} from external 
disturbances and act as geodesic reference points for the gravitational wave detection. Laser beams are exchanged between the \acp{SC} across the \SI{2.5}{Gm} arms of the constellation, tracking the distance 
variations between the \acp{TM}. Due to orbital dynamics, the frequencies of the inter-\ac{SC} lasers will be subject to Doppler shifts in the \si{\mega\hertz} band, such that the interferometers onboard \ac{LISA} will detect heterodyne frequencies with a 
 bandwidth of about $\SIrange{5}{25}{MHz}$. Distance fluctuations between the spacecraft and the \acp{TM} housed within them will be encoded as phase fluctuations in these \si{\mega\hertz} beatnotes, which the \ac{LISA} phasemeters will be able to resolve with $\mu$-cycle precision, corresponding to a design sensitivity of about $\SI{10}{pm/\rtHz}$.

This ultra precise measurement will enable \ac{LISA} to simultaneously detect and characterize tens of thousands of gravitational-wave sources, potentially answering many open questions in astrophysics, cosmology and fundamental physics.

The 
precursor mission \ac{LPF} already demonstrated the feasibility of many parts of the system 
\cite{LPF2016_Prime, LPF_prl_2018}, including the local 
interferometry inside each \ac{SC} up to approximately  
\SI{30}{\femto\meter\per\sqrt\hertz} precision 
\cite{OMS_PRL_2021,OMS_PRD_2022}. 
The inter-satellite interferometry has been partially demonstrated with the 
GRACE-FO mission \cite{grace_fo_lri}.
 
However, \ac{LISA} presents a number of unique technical challenges. Contrary to GRACE-FO, the raw readout of the inter-satellite interferometers of \ac{LISA} will be completely overwhelmed by laser frequency noise, which does not immediately cancel, due to the time-varying and unequal arms of the constellation. Instead, \ac{LISA} will make use of post processing techniques such as \acf{TDI}, in which multiple interferometric readouts are combined with the appropriate delays  to suppress the dominant noise sources, such as laser frequency noise \cite{TDI_tinto2000}. These techniques, together with strict requirements that are placed on the 
subsystems and lasers, will ensure that \ac{LISA} reaches its sensitivity goal. Different noises propagate 
through \ac{TDI} with various transfer functions \cite{tdi_noiseTF2022}, depending on their  
characteristics, such that evaluating the final performance of \ac{LISA} requires detailed studies for all performance relevant noise sources.

One of these noises is laser \acf{RIN}, which is typically described
by the laser power fluctuations relative to its average power. Since it is 
a property of each laser, it propagates through the constellation into the 
various interferometers and generates 
additive power noise to the time-varying beat signals on every 
\ac{PD}. This noise couples inevitably to the phase readout at around the 
heterodyne frequency (so called ``1f-RIN'') and its first harmonic 
(``2f-RIN'') \cite{wissel_rin2022}. We show in 
this article that it is one of the dominating metrology noise sources 
(after removal of laser frequency noise) if not carefully 
controlled. 
1f-RIN is typically the biggest 
contributor, since the resulting phase noise is scaled by the ratio of the 
beam powers, which are, for \ac{LISA}, fairly large (magnification in the 
long-arm interferometers by about 7 orders of magnitude).
As such, its impact on the sensitivity has to be understood and 
mitigated. 

We draw on the lessons learned from the \ac{LPF} mission that has also been 
used 
to study the 
effect of \ac{RIN} in a space-based heterodyne interferometer.
In this article we describe the coupling in 
the context of the \ac{LISA} mission architecture, taking into account the 
constellation characteristics, possible correlations, the optical parameters 
and the effects of \ac{LPF}-comparable mitigation schemes. 

Further, we study the impact of \ac{TDI} on the \ac{RIN} phase error by 
means of simulation and compare it with analytical expectations. We find 
that the coupling exhibits performance 
characteristics similar to that of an uncorrelated sub--$\SI{1}{pm/\rtHz}$ 
noise, assuming reasonable implementation of the mitigation strategies.

\section{Mission characteristics with respect to \ac{RIN}}
In this section, we give an overview of important mission aspects 
that have an effect on the \ac{RIN}-to-phase coupling. The theory of the 
coupling itself will be discussed in later sections.

\begin{figure}
    \includegraphics[width=0.5\textwidth]{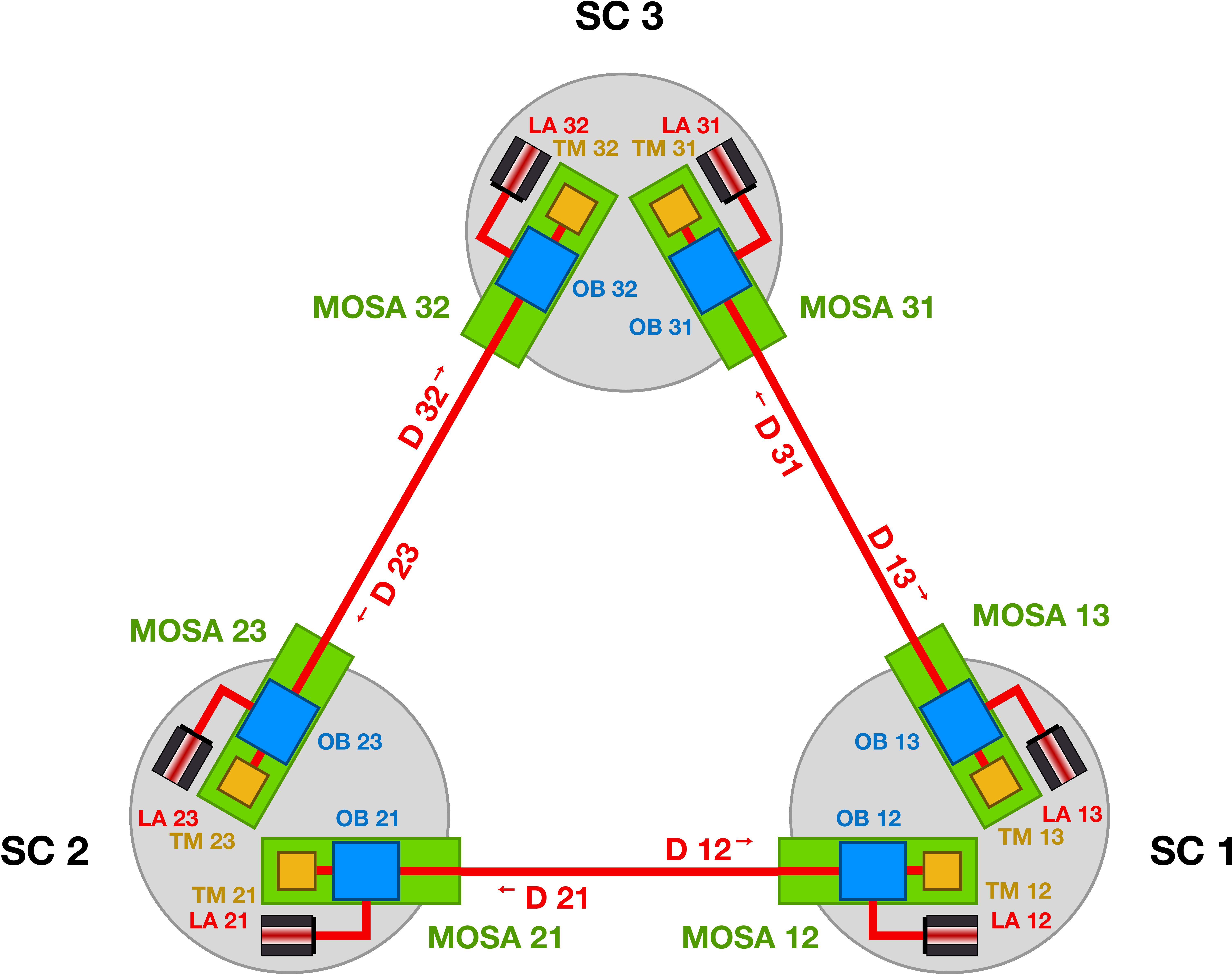}
    \caption{\label{fig:indexing} Simplified overview of the \ac{LISA} 
    triangular constellation with the naming conventions as used in this 
    article. Delays are denoted $D_{ij}$; Optical Benches $\text{OB}_{ij}$; lasers 
    are called 
    $\text{LA}_{ij}$. Reprint from 
    \cite{Hartwig:2020tdu}.}
\end{figure}

In \cref{fig:indexing}, a schematic of the constellation with the commonly-used nomenclature is shown.
The main measurement is the ``virtual'' \ac{TM}-to-\ac{TM} measurement along 
one \ac{LISA} arm. For technical reasons (e.g., beam divergence over 
millions of kilometers, very weak beam powers, 
straylight effects and optical design), no direct 
\ac{TM}-to-\ac{TM} 
measurements are possible. Therefore, we use the ``split-interferometry'' setup, in which 3 optical measurements are combined to reconstruct the desired quantity: the local \ac{TM}-to-local \ac{SC} measurement, the local \ac{SC}-to-distant \ac{SC} measurement, and the distant \ac{SC}-to-distant \ac{TM} measurement.

The total single link \ac{TM}-to-\ac{TM} metrology noise is 
considered to be below 
$\SI{10}{pm/\rtHz}$,
\begin{align}
    S_\text{IFO}^{1/2} &\le \SI{10}{\frac{pm}{\rtHz}} 
    \cdot\sqrt{f_R},\\
    f_R &= 1+\left(\frac{\SI{2}{mHz}}{f}\right)^4,
\end{align}
with $f_R$ as a factor allowing for a relaxation towards lower frequencies, 
where acceleration 
noise becomes dominant and testing is difficult \cite{LISAL3}.

\begin{figure}
    \includegraphics[width=0.5\textwidth]{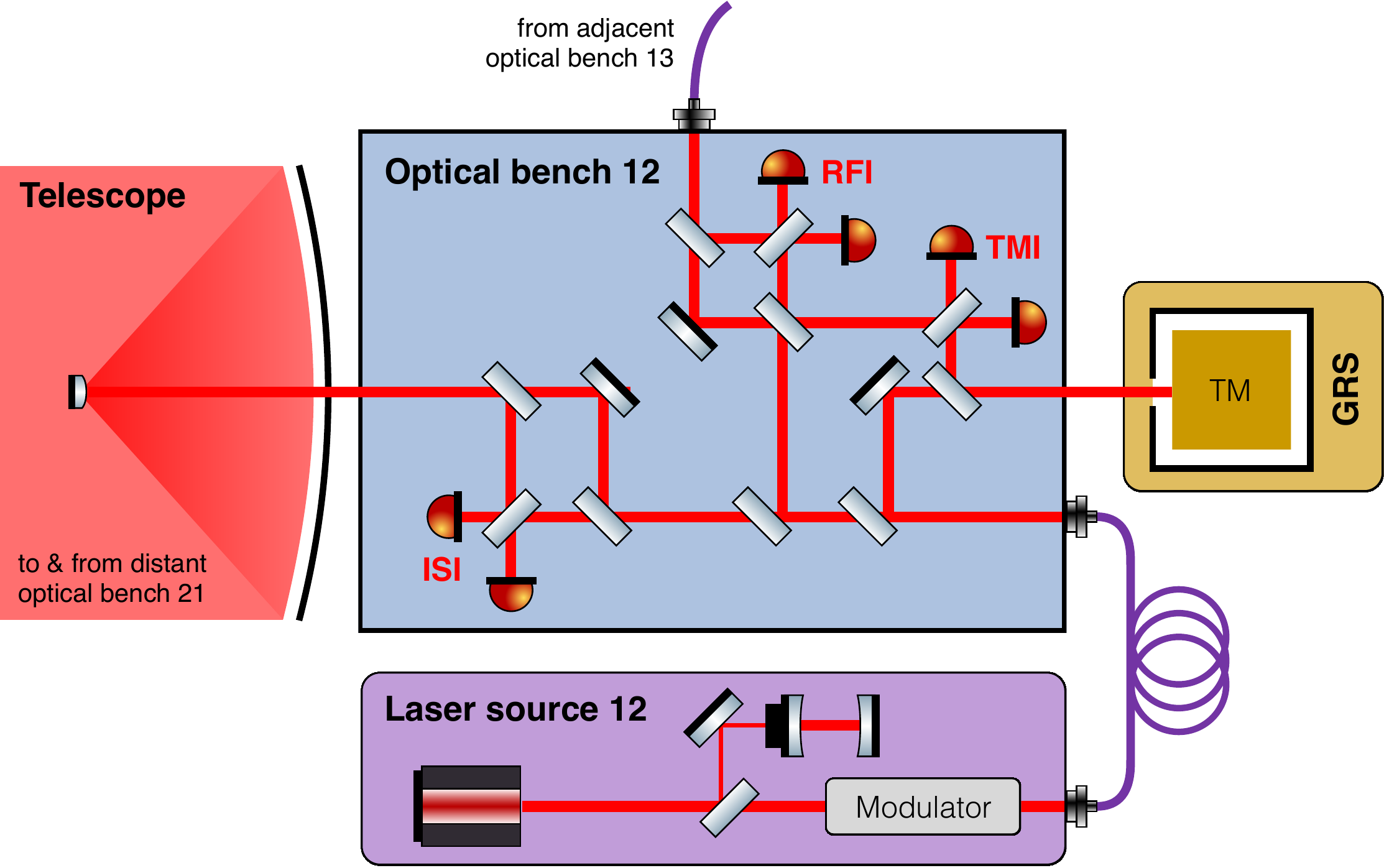}
    \caption{\label{fig:optical-design} Schematic of the optical 
    interferometry on one \ac{LISA} \ac{MOSA} from \cref{fig:indexing}. A 
    telescope collects the 
    light from the distant \ac{SC} and interferes it with the local beam. 
    The local laser is also interfered with the laser from the adjacent 
    \ac{MOSA} on the same \ac{SC} in the local interferometers (\acp{TMI} 
    and \acp{RFI}). The 
    \acf{GRS} controls the \ac{TM} relative to the 
    \ac{SC} in the suspended degrees of freedom.}
\end{figure}

The lasers have an output power of $\SI{2}{W}$ at  
$\SI{1064}{nm}$~\cite{LISAL3}, and are stabilized on a cavity.
A total of six lasers are 
powering 18 interferometers, and 
enable the \ac{TM}-to-\ac{TM} measurement by linear combinations. Per 
\ac{SC}, there are two 
\acfp{MOSA}, each attached to a laser source 
(named ``$\text{LA}_{ij}$''). They host three interferometers:
\begin{itemize}
    \item one inter-\ac{SC} interferometer (ISI) containing the \ac{GW} 
    signals,
    \item one \ac{TM}-to-\ac{SC} interferometer (TMI), used to monitor the reference points in this split interferometry setup, and
    \item one reference interferometer (RFI), used for laser locking and 
    reduction of common noise.
\end{itemize}

Between the two \ac{SC} of one \ac{LISA} arm, there are two symmetric laser 
links. Due to divergence of the Gaussian output beam, the laser power
reduces drastically over the $\SI{2.5}{Gm}$ propagation distance to a few 
hundreds \si{\pico\watt} at the receiving \ac{SC}; it is then interfered 
with a local \si{\milli\watt} beam.
The laser beams will carry additional modulation sidebands used for clock 
synchronization, ranging information and data transfer, which further reduce 
the available 
power in the main carrier-to-carrier beat 
signal to about $\SI{81}{\%}$ \cite{LISAL3,hartwig_thesis}.

The two adjacent \acp{MOSA} exchange their laser light via 
fiber backlinks, see \cref{fig:optical-design}. To reduce backscatter, the 
powers guided into the fibers are also relatively small (in the order of 
\si{\milli\watt} to \si{\nano\watt}) and are interfered with beams a few 
orders of magnitude stronger, 
such that the beam power ratio in any interferometer is far from unity. 
By design, the laser beam properties in each interferometer are 
different. Even comparing the two local interferometers between each 
\ac{MOSA} on a single \ac{SC}, 
which receive beams from the same laser source and share the same absolute 
beat frequency, shows that they have their power ratios inverted due to the 
fiber transfer; 
thus the local \ac{SC} scaling of the \ac{RIN}-to-phase couplings (that 
depend on the power ratios for 1f-RIN) are not completely identical, as will 
be considered later in this article.

In every interferometer, two beams interfere at a recombination beamsplitter 
and \acp{PD} measure their impinging time-varying power.
The two output ports of each of these beamsplitters are used to apply 
balanced detection to the 
(naturally $\pi$-shifted)
signals, which allows us to subtract both ports to 
reduce noises like 1f-RIN, while maintaining the signal 
information \cite{wissel_rin2022}.

\begin{figure}
    \includegraphics[width=0.5\textwidth]{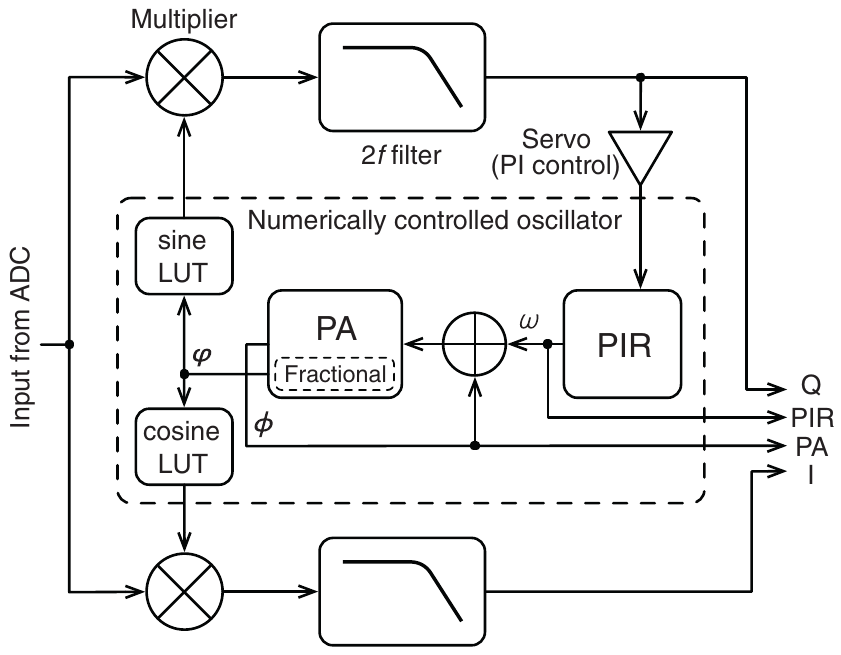}
    \caption{\label{fig:dpll} Schematic of a \ac{DPLL} in 
        \ac{LISA}. The input time series from the \ac{ADC} is mixed with the 
        sine from an \ac{NCO}, which represents a closed control loop with 
        the 
        down-mixed instantaneous frequency as its error signal. The 
        \ac{DPLL} is 
        able to follow the input frequency (even for time-varying heterodyne 
        frequencies) within its bandwidth and accumulates the total phase of 
        the 
        input, which is the desired phase measurement. PA phase 
        accumulator; PIR 
        phase-increment register; LUT look-up table; PI 
        proportional-integral. Reprint from 
        \cite{heinzel_newDWS2020}.}
\end{figure}

The phase measurement is performed by dedicated \acp{DPLL}  
\cite{Gerberding_2013,heinzel_newDWS2020}, as depicted in 
\cref{fig:dpll}. The loops are able to 
track the time-varying beatnote over many MHz and 
measure the phase with $\mu$-cycle precision. It resembles a typical 
I/Q-demodulation scheme, but is all performed digitally and uses a control 
loop on the $Q$ quadrature as an error signal for an \ac{NCO} to drive the 
mixing process.

\begin{figure}
    \includegraphics[width=0.5\textwidth]{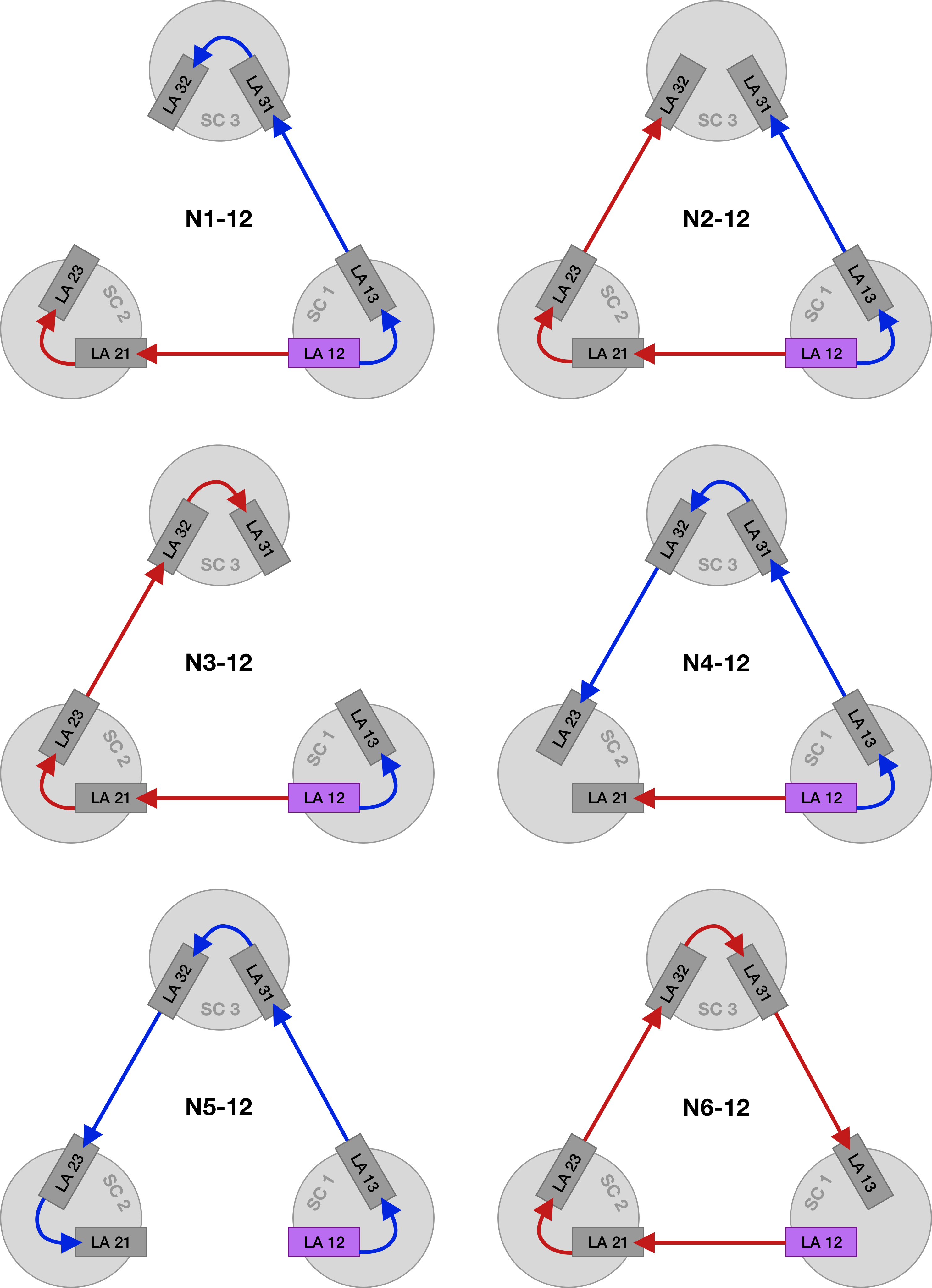}
    \caption{\label{fig:locking-configurations} Schematic of the different 
    locking configurations, here with laser $\text{LA}_{12}$ as the primary 
    laser. Reprint from \cite{hartwig_thesis}.}
\end{figure}

Since each laser is involved in 6 interferometers, they can possibly add 
correlated noise in those interferometers. However, the \ac{RIN}-to-phase 
coupling 
depends on the absolute beat frequency (and its harmonic), which 
means that \acp{RIN} at different heterodyne frequencies in different 
interferometers can be considered 
uncorrelated if the beatnote frequencies are reasonably well separated, i.e., by more than the 
measurement band of a few \si{\hertz}; this holds even if the same laser is involved. \ac{RIN} 
from different laser sources is always considered uncorrelated, especially at MHz frequencies. This means that \ac{RIN} in the 
\acp{TMI} and \acp{RFI} on the same \ac{SC} is correlated, while the 
\acp{ISI} may not have correlations.
Furthermore, the \acp{ISI} are subject to orbital Doppler 
shifts (in the MHz range)
and thus their heterodyne frequencies vary. The expected shifts can be 
calculated beforehand and are used to enable and optimize the interferometry 
and 
detection 
process. The absolute beat frequencies are technically restricted to a range 
of 
approximately $\SIrange{5}{25}{MHz}$ via an   
(offset) frequency locking scheme of the lasers. This results in a 
configuration where one primary laser is locked to a 
cavity, while 
the other five lasers are locked (with \si{\mega\hertz} offsets) to 
the  
primary laser. The required offset frequencies are calculated on ground, 
yielding a so-called frequency 
plan. Various possible locking topologies (with $\text{LA}_{12}$ as the primary laser) have been 
identified 
\cite{fplan_tn, 
hartwig_thesis}, as shown in \cref{fig:locking-configurations}.

The locking also inevitably imprints any noise (and signal) information  
of the interferometers 
used for locking onto the locked lasers. Therefore, \ac{RIN}-induced phase 
noise will also be added to the locked laser 
and propagates through the constellation into all six interferometers 
of that laser. 
Furthermore, the next laser that locks on the first locked laser will 
continue to carry this noise and thus have locally increased phase noise.

\begin{figure}
    \includegraphics[width=0.5\textwidth]{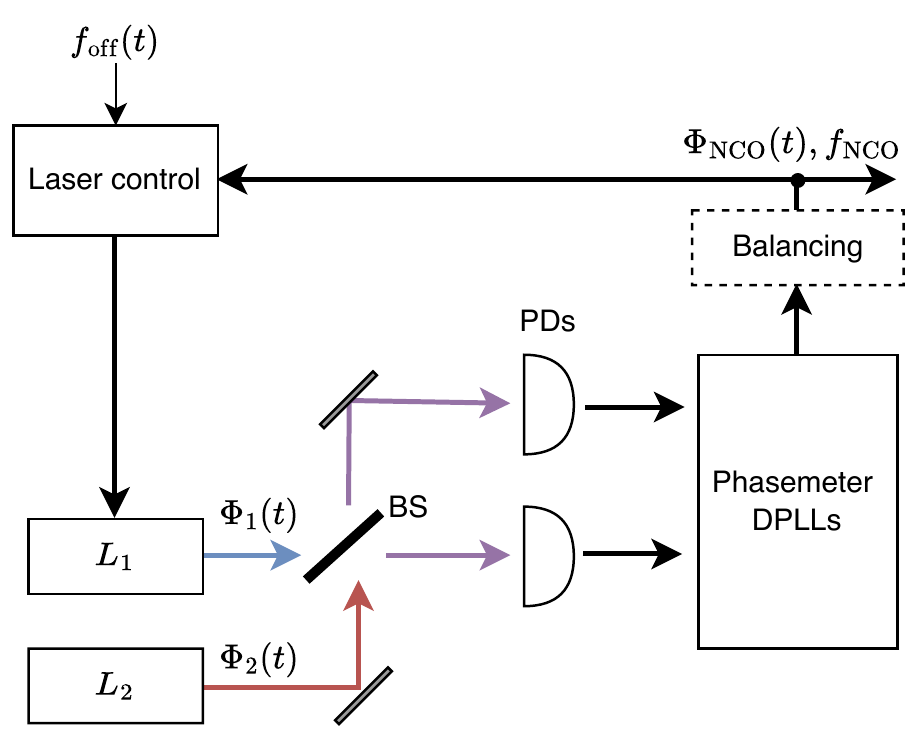}
    \caption{\label{fig:offset_freq_lock} Schematic of the outer offset 
    frequency 
        locking control loop with \acp{DPLL}. This represents the simplified 
        view 
        of one \ac{LISA} interferometer, which is used for locking laser $L_1$ 
        to 
        laser $L_2$ with a typical bandwidth of multiple $\si{kHz}$. The 
        two lasers with phase evolution $\Phi_i(t)$ are brought to 
        interference via 
        the \ac{BS}, detected on the \acp{PD} (in reality redundant 
        \acp{QPD}) 
        and measured in the phasemeter. 
        The instantenous (balanced) frequencies of the \ac{DPLL} are used to 
        offset 
        lock the lasers according to the predetermined frequency 
        planning.}
\end{figure}

\Cref{fig:offset_freq_lock} shows one possible schematic for the local 
laser control loop used for locking. Here, we assume that the error signal 
of the control loop has been balanced between the two interferometer output 
ports (current baseline), which propagates only a reduced amount of 
phase noise ``echos'' through the constellation.

We calculate and simulate these effects in this paper. 
Luckily, \ac{TDI} strongly suppresses any laser phase fluctuations in 
post-processing, including echoes from the locking control loops, such 
that the final \ac{TDI} variables are unaffected by the choice of the 
locking topology \cite{hartwig_thesis,tdi_noiseTF2022}. 

\section{RIN coupling in LISA}

The \ac{RIN} $n = P(t)/\langle P(t) \rangle$ of 
any laser power $P(t)$, usually expressed in \ac{ASD} units of 
$1/\rtHz$, causes phase noise in the interferometric readout via 
three distinct coupling channels \cite{wissel_rin2022,LPF_prl_2018}. First, ``DC-RIN'' inside the measurement band at low frequencies 
causes slow intensity fluctuations that lead to radiation pressure on the 
\acp{SC} (negligible) and \acp{TM} (not negligible). 
This drives 
the low-frequency \ac{RIN} requirements and has an 
assumed level of $\SI{100}{ppm\per\rtHz}$ at 
$\SI{0.1}{mHz}$.
It gives a small contribution to the \ac{TM} acceleration of about 
$\SI{0.35}{fm/s^2/\rtHz}$, out of a total acceleration noise allocation of roughly $\SI{10}{fm/s^2/\rtHz}$ at $\SI{0.1}{mHz}$ \cite{LISAL3}.

Secondly, 1f-RIN from around the 
heterodyne 
frequency causes 
additive phase noise to the main signal on the \acp{PD}.

Finally, 2f-RIN 
from around twice the heterodyne frequency is optically down-mixed and also 
produces additive noise (to first order) at the heterodyne frequency. 

We 
focus in this article on the latter two mechanisms. They do not cause 
any direct force noise, and therefore do not generate any real \ac{TM} motion signal, but cause an
additive small-vector readout noise instead.

Since the \ac{RIN} is a property of the laser beams, it appears correlated 
on all \ac{QPD} segments of a single diode. In this paper, we use the terms \ac{QPD} and \ac{PD} interchangeably, as it has no impact on our description of the
\ac{RIN} coupling in the longitudinal degree of freedom.

\subsection{Theory of \ac{DPLL} readout}
In the next sections, we derive the \ac{RIN}-to-phase coupling equations using the \ac{LISA}-specific \ac{DPLL} readout architecture.

The \ac{DPLL} depicted in \cref{fig:dpll} is a control loop that uses the 
error 
signal, $Q$, to adjust the total phase and frequency of an \ac{NCO}. 
Contrary 
to a 
simple I/Q demodulation, the phase readout is not 
directly given by combining the I/Q channels, but instead available by 
digitally reading out the NCO registers. 

When the loop is closed, and if we assume it to work perfectly with infinite 
gain in the measurement 
band, the error signal will be exactly equal to zero. The error signal is 
produced by multiplying the incoming signal (modeled as a cosine) by a sine 
that is perfectly in phase.

We consider an input signal that has a strong main beatnote plus small 
additive disturbing terms, representing   
small-vector \ac{RIN}. We write
\begin{equation}
    V(t) = A \cos(\Phi(t)) + n(t),
\end{equation}
where we assume $A$ to be constant, and $\abs{n(t)} \ll A$. Here, $\dot 
\Phi(t)$ is typically in the order of MHz. $n(t)$ represents our 
different 
\ac{RIN} terms, but could in principle also be any other additive noise.

We assume that the phase error caused by the disturbance $n(t)$ is small, of 
order $\ll 1$ expressed in cycles or radian. Therefore, we consider the 
\ac{NCO} signal used in the lock to closely follow the main beatnote, and 
model it as 
\begin{equation}
    U(t) = \sin(\Phi(t) + \varphi(t)),
\end{equation}
where $\varphi(t)\ll 1$ accounts for the phase readout errors due to the 
disturbance $n(t)$. The total phase 
\begin{equation}
    \Phi_\text{NCO}(t) = \Phi(t) + \varphi(t)    
\end{equation}
represents our phase readout and is available from the phasemeter 
phase accumulator.

The error signal is then computed by mixing the \ac{NCO} signal with the 
input signal,
\begin{equation}
    Q(t) = \langle V(t) U(t) \rangle,
\end{equation}
where $\langle \cdot \rangle$ denotes a low-pass filter removing 
frequency 
content far away from DC.  We assume that this filter is a linear operation, 
in the sense that $\langle a X + b Y\rangle = a \langle  X \rangle + b 
\langle  Y\rangle $. A typical example for such a filter is a moving average.

The loop will adjust the phase of the \ac{NCO} to drive the error signal to 
zero. 
This means we can model how the disturbance $n(t)$ affects the output of the 
\ac{DPLL} for the closed loop by solving the equation $Q \equiv 0$ for 
$\varphi$; i.e., by finding the \ac{NCO} phase for which the error signal 
vanishes. 

\subsection{Phase readout}




Combining the previous equations, we can write
\begin{align} 
    \begin{split}
    Q ={}&  \langle A 
\cos(\Phi(t)) \sin(\Phi(t) + \varphi(t)) \rangle   \\ 
& + \langle n(t) 
\sin(\Phi(t) + \varphi(t)) \rangle .
    \end{split}
\label{eqn:pll_Q_with_phase_noise}
\end{align}

Using trigonometric identities and that $\varphi(t) \ll 1$, the first term on the right-hand side
\begin{equation}
    \langle A 
\cos(\Phi(t)) \sin(\Phi(t) + \varphi(t)) \rangle \approx \frac{A}{2} \varphi(t).
\end{equation}
To treat the other term, we first expand to first order in $\varphi$ and then neglect the second-order term containing $\varphi(t) n(t)$, yielding
\begin{equation}
    \langle n(t) \sin(\Phi(t) + \varphi(t)) \rangle \approx \langle n(t) \sin(\Phi(t))\rangle.
    \label{eqn:pll_Q_with_phase_noise_simplified}
\end{equation}

Using this in 
\cref{eqn:pll_Q_with_phase_noise}, with the locking condition $Q \equiv 0 $, 
gives the phase error induced by $n(t)$,
\begin{equation}
    \varphi(t) \approx - \frac{2}{A} \langle n(t) \sin(\Phi(t))\rangle,
    \label{eq:early-expansion}
\end{equation}
and the total phase readout will be given as
\begin{equation}
    \Phi_\text{NCO}(t) \approx \Phi(t) - \frac{2}{A} \langle n(t) 
    \sin(\Phi(t))\rangle.
    \label{eq:early-expansion-nco}
\end{equation}
This means that, to first order, the disturbance is simply mixed with the main 
beatnote and scaled by the reciprocal beatnote amplitude.


\subsection{Scaling for \ac{RIN}}

We now need to apply
\cref{eq:early-expansion-nco} to the typical 
photodiode detection equations adapted for heterodyne interferometry. They provide 
the scaling factors for $A$ and $n(t)$ that describe the \ac{RIN} coupling 
correctly. The equations are derived in 
\cite{wissel_rin2022}, and are given here with the relevant \ac{RIN} terms 
only. Note that any input DC contributions are neglected here.
We use the equations adapted for \ac{LISA} to describe the interferometer 
output ports A and B of 
the recombination beamsplitter (with 
amplitude transmission and reflection coefficients $\tau, \rho$, average 
beam powers $P_i$, heterodyne 
efficiency $\eta_\text{het}$, and RIN $n_{m}, n_{r}$ for a general 
measurement and 
general reference beam $m, r$ and a certain signal power in the carrier of 
$\epsilon_\text{carrier}$). They yield, for the measured powers per output port,

    \begin{align}\label{e:rin_detection_eqn}
    \begin{split}
        P_\text{A} &= \underbrace{\rho^2 P_\text{m} n_\text{m} + \tau^2 
        P_\text{r} 
            n_\text{r}}_\text{1f-RIN, port A} 
        \\ 
        & \underbrace{+(n_\text{m}+n_\text{r}) \rho \tau 
        \epsilon_\text{carrier}
        \sqrt{\eta_\text{het} P_\text{m} 
                P_\text{r}} \cos(\Phi(t))}_\text{2f-RIN, port A}    
        \\ 
        &\underbrace{+ 2 \rho \tau \epsilon_\text{carrier} 
        \sqrt{\eta_\text{het} P_\text{m} 
        P_\text{r}} \cos(\Phi(t))}_\text{Signal, port A},  
    \end{split}
    \end{align}
and
    \begin{align}
        \begin{split}
        P_\text{B} &= \underbrace{\tau^2 P_\text{m} n_\text{m} + \rho^2 
        P_\text{r} 
           n_\text{r}}_\text{1f-RIN, port 
            B} 
        \\ 
        & \underbrace{-(n_\text{m}+n_\text{r}) \rho \tau 
        \epsilon_\text{carrier}
        \sqrt{\eta_\text{het} P_\text{m} 
                P_\text{r}} \cos(\Phi(t))}_\text{2f-RIN, port B} 
        \\ 
        &\underbrace{- 2 \rho \tau \epsilon_\text{carrier} 
        \sqrt{\eta_\text{het} P_\text{m} 
        P_\text{r}} \cos(\Phi(t))}_\text{Signal, port B}.
    \end{split}
    \end{align}
Here, we already see that balanced detection of the form $(P_A-P_B)/2$ is 
able to suppress 1f-RIN, since it appears with the same sign in both ports. 
However, 2f-RIN appears with opposite signs in the two ports, identical to the main signal, and therefore cannot be suppressed by balanced detection.

From these equations we can model the input signal 
to the \ac{DPLL} using
\begin{equation}
    V(t) = \pm A \cos(\Phi(t)) + a_i n_i(t) \pm \frac{A}{2} n_i(t) 
    \cos(\Phi(t)),
\end{equation}
with the scale factor $A = 2  \rho \tau \epsilon_\text{carrier}
\sqrt{\eta_\text{het} P_\text{m} 
P_\text{r}}$, while $a_i$ represents the scale factor for one of the 1f-RIN 
terms and $n_i(t)$ the \ac{RIN} of one of the beams. 
The $\pm$ 
encodes output 
port A or B. Since the \ac{RIN}  
between the two beams is uncorrelated (as well as 1f- and 2f-RIN per beam), 
we can calculate their resulting phase noise independently. In terms of 
spectral densities, one can build their quadratic sum for the total 
phase noise afterwards.

We assume a relative power 
stability of the lasers of around $\SI{3e-8}{\hertz^{-1/2}}$ in the 
relevant bandwidth, such that the resulting phase noise will be small, 
$\varphi(t) \ll 1$. This allows to use the previous result of 
\cref{eq:early-expansion}.

Therefore, we insert $n(t) =  a_i n_i(t) \pm \frac{A}{2} n_i(t) 
\cos(\Phi(t))$ 
into \cref{eq:early-expansion}, which gives for the resulting phase noise (generated by one laser beam $i = {m,r}$),
\begin{equation}
    \varphi_i(t) = \underbrace{-\frac{2a_i}{A}\langle n_i(t) 
    \sin(\Phi(t))\rangle}_\text{1f-RIN phasenoise}  
    \underbrace{\mp\frac{1}{2}\langle n_i(t) \sin(2\Phi(t)) 
    \rangle}_\text{2f-RIN phasenoise}.
\end{equation}
We find that the noise $n(t)$ 
appears mixed both with $\sin(\Phi(t))$ as well as 
$\sin(2\Phi(t))$. This implies that noise around $\dot \Phi(t)$ and $2\dot 
\Phi(t)$ are down-converted to the phasemeter base band and couple into the 
phase accumulator. Due to the down- and up-conversion process of the mixing, 
only half of that noise power is actually contributing to $\varphi(t)$, 
since the other half is filtered out. We also see that the 2f-RIN is 
independent of the signal amplitude or average beampowers.

The total \ac{RIN}-to-phase coupling must contain the \ac{RIN} from both 
beams, 
\begin{equation}
    \varphi_\text{tot}(t) = \varphi_m(t) + \varphi_r(t),
\end{equation}
with the corresponding coupling factors $a_m = \rho^2 P_m$, $a_r = \tau^2 
P_r$ 
in port A and $a_m = \tau^2 P_m$, $a_r = \rho^2 P_r$ in port B.

After balanced detection, these coupling factors become
\begin{align}
    A &=  2 \rho \tau\epsilon_\text{carrier}  \sqrt{\eta_\text{het} 
    P_\text{m} 
        P_\text{r}}, \\
    a_m &= (\rho^2 P_m - \tau^2 P_m)/2,\label{eq:balancing-m} \\
    a_r &= (\tau^2 P_r - \rho^2 P_r)/2.\label{eq:balancing-r}
\end{align}

These mixing equations are used 
in the simulation results presented in later sections to carry over any phase-correlation 
information correctly. They also agree with the results derived in 
\cite{wissel_rin2022} for a small-vector noise approach, which is slightly 
less general.

Each interferometer (\ac{ISI}, \ac{TMI}, \ac{RFI}) will carry such a phase 
error $\varphi_\text{tot}$ (here named by interferometer and usually scaled 
by 
$\lambda/(2\pi)$),
\begin{align}
    \varphi_\text{ISI}(t) &= \varphi_{\text{ISI},m}(t) + 
    \varphi_{\text{ISI},r}(t), \\
    \varphi_\text{TMI}(t) &= \varphi_{\text{TMI},m}(t) + 
    \varphi_{\text{TMI},r}(t), \\
    \varphi_\text{RFI}(t) &= \varphi_{\text{RFI},m}(t) + 
    \varphi_{\text{RFI},r}(t),
\end{align}
where the different 1f-RIN amplitudes and total phases depend on the 
different interferometer optical settings (see \cref{t:optical_params}).
\begin{table*}
    \begin{center}
        \begin{tabularx}{1\linewidth}{llX}
            \toprule
            \textbf{Parameter} & \textbf{Value} & \textbf{Description}  \\ 
            \midrule
            $\lambda$ & \SI{1064}{nm} & Laser wavelength \\
            $f_{\text{het}}$ & $\SIrange{5}{25}{MHz} $ & Heterodyne 
            frequency\\
            $n_\text{1f, 2f}$ & $ 
            \SI{3e-8}{\frac{1}{\sqrt{Hz}}}$ & 
            Maximum (white) RIN \ac{ASD} in the band \SIrange{5}{50}{MHz}  \\
            $\tau^2, \rho^2$ & $0.5$ & Beamsplitter (in power when squared 
            as 
            given) transmission and 
            reflection coefficients \\
            $b$ & $[\SI{0.9}{}, \SI{0}{}, \SI{1}{}] $ & 
            Balancing efficiency, i.e. matching of the 1f-RIN amplitudes in 
            the interferometer ports  \\
            $\eta_{\text{het,ISI}}$ & $0.75$ & Heterodyne efficiency from 
            the overlap integral in a 
            long-arm (\ac{ISI}) interferometer  \\ 
            $\eta_{\text{het,TMI}}$ & $0.82$ &Heterodyne efficiency from the 
            overlap integral in a 
            \ac{TM} (\ac{TMI}) interferometer \\
            $\eta_{\text{het,RFI}}$ & $0.82$ & Heterodyne efficiency from 
            the overlap integral in a reference (\ac{RFI}) 
            interferometer \\
            $\epsilon_\text{carrier}$ & \SI{0.81}{} & Portion of power in 
            the carrier of the beams \\
            $P_{\text{ISI},1}$ & $\SI{350}{pW}$ &  Mean power of the 
            remote laser in the long-arm interferometer (from distant 
            \ac{SC}) \\ 
            $P_{\text{ISI},2}$ & $\SI{1}{mW}$ & Mean power of the 
            local laser in 
            the long-arm interferometer \\ 
            
            $P_{\text{TMI},1}$ & $\SI{500}{nW}$ & Mean power of the 
            adjacent laser in the \ac{TM} interferometer (from adjacent 
            bench)  \\ 
            $P_{\text{TMI},2}$ & $\SI{500}{\mu W}$&  Mean power of the 
            local laser in the \ac{TM} interferometer  \\ 
            
            $P_{\text{RFI},1}$ &$\SI{500}{nW}$ &  Mean power of the 
            adjacent laser in the reference interferometer (from adjacent 
            bench)  \\   \\ 
            $P_{\text{RFI},2}$ & $\SI{1}{mW}$ & Mean power of the 
            local laser in the reference interferometer  \\ 
            
            $dx_{\text{SC}}$ & \SI{10}{nm/\sqrt{Hz}} & Residual 
            translational jitter of the \ac{SC}, 
            with respect to inertial space.  When used as a residual path 
            offset in the 
            equations, we calculate a \ac{RMS} value by integrating over a 
            frequency band 
            from 
            \SIrange{0}{1}{Hz} as in $\sqrt{(\SI{10}{nm/\sqrt{Hz}})^2 \cdot 
                \SI{1}{Hz}} = 
            \SI{10}{nm}$. In \ac{LPF} we measured a \ac{RMS} in $x_1$ of 
            about 
            \SI{2}{nm}, 
            and a 
            peak-to-peak difference of about \SI{10}{nm}   \\
            $\phi_\text{ISI}$ & $dx_{\text{SC}}$ & Without noises or signals 
            the measured phase due to minimal residual
            translational jitter in an \ac{ISI} \\
            $\phi_\text{TMI}$ & $ dx_{\text{SC}}$, $[-2\pi,2\pi]$  & Limit 
            of \ac{TMI} interferometer set-point due to residual jitter or 
            \ac{TM} guidance injection \\
            $\phi_\text{RFI}$ & $\SI{0}{rad}$ &  \ac{RFI} interferometer 
            phase offset \\
            \bottomrule
        \end{tabularx}
    \end{center}
    \caption{Parameters of the optical chain with special relevance for the 
        \ac{RIN} to phase coupling. Where multiple numbers are stated they 
        correspond to the different simulations performed. The values are 
        estimates from current design studies, and are subject to minor 
        changes.}
    \label{t:optical_params}
\end{table*}

Note that for our simulations, we do not use 
\cref{eq:balancing-m,eq:balancing-r}, but instead 
model imperfections in the balanced detection by artificially introducing a 
balancing efficiency, $0\leq b\leq1$, and then model the residual 1f-RIN 
terms as
\begin{equation}
    \varphi_{\text{1f},i}(t) = -\frac{2a_i}{A}\langle n(t) 
    \sin(\Phi(t))\rangle \cdot (1-b),
\end{equation}
while still using perfect 50/50 beamsplitters in the simulation.

\subsection{Simplified phase noise equations without correlations}
If only the maximum or \ac{RMS} \ac{RIN}-to-phase coupling is required (for 
example for the noise level in only one interferometer), one 
can simplify the equations above by dropping the phase information in the 
mixing process.
This ignores correlation properties of \ac{RIN} but 
still gives the right level of phase noise per individual interferometer. We 
use the effect of the filter to select two independent noise series at the 
in-band sampling frequency. These two time-series $n_{1f}(t), n_{2f}(t)$ 
have to be scaled 
due to the mixing and filtering process and represent \ac{RIN} from around 
$\dot 
\Phi$ (1f-RIN) and $ 2\dot\Phi$ (2f-RIN). The scale factors arise from 
simplifying $\langle 
n(t) \sin(\Phi(t))\rangle$, which has an \ac{ASD} of approximately 
$\frac{1}{\sqrt{2}} \tilde{n}$, with  $\tilde{n}$ as the  \ac{ASD} value of 
$n(t)$.

Therefore, if we 
want to 
replace the mixing and filtering process with an in-band, downsampled 
version of 
the $n(t)$ noise that has the correct scaling, we can use 
\begin{align}
    \langle n(t) \sin(\Phi(t))\rangle &\approx \frac{1}{\sqrt{2}} 
    n_{1f}(t),
    \\
    \langle n(t) \sin(2\Phi(t))\rangle &\approx \frac{1}{\sqrt{2}},
    n_{2f}(t)
\end{align}
for the 1f and 2f-RIN terms, respectively. We typically assume 
$\tilde{n}_{1f} = \tilde{n}_{2f} = \SI{3e-8}{\hertz^{-1/2}}$.
In total, that gives for one beam
\begin{equation}
    \varphi_i(t) \approx -\frac{2a_i}{A}\frac{n_{1f}(t)}{\sqrt{2}} - 
    \frac{1}{2}\frac{n_{2f}(t)}{\sqrt{2}}.
\end{equation}
The total \ac{RIN}-induced phase noise from two beams would then be the same 
sum as before, $\varphi_\text{tot}(t) = \varphi_m(t) + \varphi_r(t)$, but 
now expressed simply by four uncorrelated noise time series with low 
sampling 
frequency and corresponding standard deviation.
This can be used to set upper boundaries per interferometer level or  
for the locally correlated measurements (by using correlated time series 
for upper-boundary estimates).

\subsection{Laser locking}
The required control for laser locking adds another outer loop that uses the 
instantaneous frequency measured by the inner \ac{DPLL}, as depicted in 
\cref{fig:offset_freq_lock}. Each laser has its own phase, $\Phi_1(t), 
\Phi_2(t)$, such that the beatnote is represented by the total difference 
phase $\Phi(t) = 
\Phi_1(t) - \Phi_2(t)$. As before, we denote by $n(t)$ an additive noise 
source, in our case \ac{RIN}.

From the locking condition of the \ac{DPLL}, \cref{eq:early-expansion-nco},  
we know that 
\begin{equation}
    \Phi_\text{NCO}(t) \approx \Phi(t) - \frac{2}{A}\langle n(t) 
    \sin(\Phi(t)) \rangle \nonumber.
\end{equation}
The outer offset frequency locking loop has the (ideal in-band) locking 
condition that
\begin{align}\label{eqn:freq_lock_condition}
    f_\text{NCO}(t) - f_\text{off}(t) &\equiv 0, \\
    \Longleftrightarrow \,\, \dot \Phi_\text{NCO}(t) - f_\text{off}(t) 
    &\equiv 0
\end{align}
where the frequency offsets $f_\text{off}(t)$ are predefined values 
calculated from the frequency planning. 

This locking is assumed to be perfect (in the measurement band, within a 
bandwidth of $\sim \SI{10}{kHz}$), such that any disturbance of the offset 
phase, for example due to \ac{RIN}, laser frequency noise, or even 
gravitational-wave signals, is added sign-inverted to the laser phase  
(compared to the phasemeter measurement). This implies that any of those 
terms
cancel in the 
respective interferometer. Note that we assume the current baseline, i.e., that the 
locking will use the balanced readout, such that 1f-RIN is already minimized 
at the input to the control loop and therefore not fully imprinted on the 
laser, and thus not canceled on the level of individual photodiodes. Since 
2f-RIN is not canceled by balanced detection, it will be fully imprinted on 
the laser, i.e., it will be fully canceled on the level of individual 
photodiodes.



We can see this easily if we look at the outer loop locking equation. In 
the phase domain, after integrating \cref{eqn:freq_lock_condition} and setting 
the integration constant to 0, we have
\begin{align}
    0 &\equiv \Phi_\text{NCO}(t) - \Phi_\text{off}(t)  \\
    \Rightarrow \Phi_\text{off}(t)  &\approx  \Phi(t) - \frac{2}{A}\langle 
    n(t) \sin(\Phi(t)) \rangle  \\
    \Longleftrightarrow \Phi_\text{off}(t) &\approx  \Phi_1(t) - \Phi_2(t) 
    \nonumber \\
    &\quad- 
    \frac{2}{A}\langle n(t) \sin(\Phi_1(t) - \Phi_2(t)) \rangle  \\
    \Rightarrow \Phi_1(t) &= \Phi_2(t) + \Phi_\text{off}(t) \nonumber \\
    &\quad + \frac{2}{A}\langle n(t) 
    \sin(\Phi_1(t) - \Phi_2(t)) \rangle .
\end{align}
We see that the equation is implicit and cannot be solved analytically  for 
$\Phi_1(t)$. We can iteratively solve it to first order,
\begin{align}
    \Phi_1(t) &= \Phi_2(t) + \Phi_\text{off}(t) \nonumber \\
    &\quad + \frac{2}{A}\langle n(t) \sin([\Phi_1(t) =\dots] 
    - \Phi_2(t)) \rangle  \\ 
    &\approx \Phi_2(t) + \Phi_\text{off}(t) + \frac{2}{A}\langle n(t) 
    \sin(\Phi_\text{off}(t)) 
    \rangle   \\
    &\approx \Phi_2(t) + 
    \Phi_\text{off}(t) + \frac{2}{A}\langle n(t) \sin(\Phi(t)) \rangle.
\end{align}
We expect that the effect of the loop 
on the input signal $\Phi_1(t)$ is very small due to $n \ll 1$, which allows 
the approximations in the last two equations for simulation purposes. It is 
assumed here that the loop will always have enough bandwidth to perfectly 
cancel the measured noise and imprint it on the laser phase, ``echoing'' it 
through the constellation.

As an example, we give the locking propagation equations for the baseline N1-12 
locking configuration, see \cref{fig:locking-configurations}; laser phase is 
locked with the phase error measured in 
the respective locking 
interferometer, and propagated with delays such that the laser phase errors 
can be written as
\begin{align}
    R_{12} &= \text{Reference}, \\
    R_{13} &= \sign(t_{13})\cdot t_{13}, \\
    R_{31} &= \sign(s_{31})\cdot s_{31} + D_{31}(R_{13}), \\
    R_{32} &= \sign(t_{32})\cdot t_{32} +  R_{31}, \\
    R_{21} &= \sign(s_{21})\cdot s_{21}, \\
    R_{23} &= \sign(t_{23})\cdot t_{23} + R_{21},
\end{align}
where $R_{ij}$ describes the \ac{RIN} phase error that each laser $\text{LA}_{ij}$ 
is 
carrying. The shorthand notation $t_{ij}$ stands for the \ac{RIN} phase 
error originating from $\text{RFI}_{ij}$ and $s_{ij}$ maps the \ac{RIN} 
phase error from $\text{ISI}_{ij}$. Delays $D_{ij}$ are defined as in 
\cref{fig:indexing} and the $\sign(\cdot)$ represents the sign of the 
beat frequency in the corresponding interferometer.

\subsection{\ac{RIN} Correlations in \ac{LISA}}
In this section, we discuss possible correlations between interferometers.

First, between the two \acp{ISI} where the same laser is involved, a 
potential  short-time 
correlation can appear if the beatnote frequencies are identical at the 
times the measurements are combined in \ac{TDI}, i.e., at multiples of the 
light travel time between the \ac{SC}. This is very unlikely to happen due 
to 
the frequency planning and the arm-breathing. 

In any case, the overall \ac{RIN} contributions in the \acp{ISI} will be 
completely dominated by the \ac{RIN} of the local beam, due to the power 
ratios $\sqrt{P_{\text{ISI},1}/P_{\text{ISI},2}} 
\cdot n_\text{1f}(t) \ll \sqrt{P_{\text{ISI},2}/P_{\text{ISI},1}} \cdot 
n_\text{1f}(t)$. Thus, any potential correlation would involve one of the 
negligible terms, and can therefore be safely ignored.

Second, as explained before, 1f-RIN 
and 
2f-RIN are several MHz apart due to the mission design, and thus can be considered as uncorrelated. However, the 
same two 
lasers interfere in the four local \acp{TMI} and \acp{RFI} on 
each \ac{SC} and therefore produce correlated \ac{RIN}. The correlation (1f- 
with 1f-RIN) is not very strong, since the power ratios are inverse in the 
local interferometers of two adjacent \acp{MOSA}. In the case of 2f-RIN, there is full 
correlation, since it does not depend on the beam powers. 

\begin{figure}
    \includegraphics[width=0.49\textwidth]{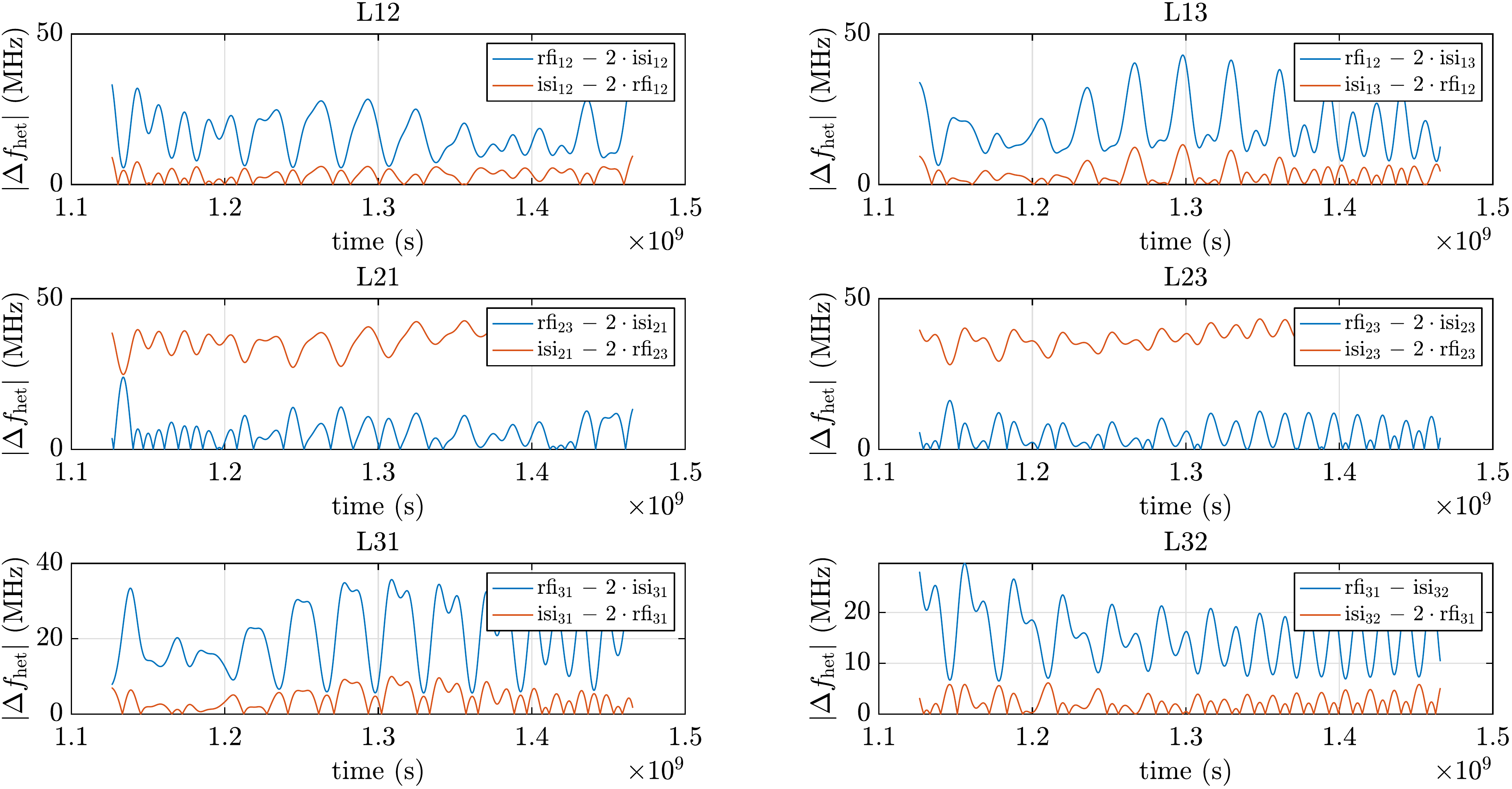}
    \caption{\label{fig:locking_correlations} Locking beatnote differences 
    between local and \ac{ISI} interferometers per laser. It shows the 
    possibility of 1f-2f-RIN  
    correlations for the baseline configuration N1-12 for more 
    than 10 years of data. We show pairwise absolute beat frequency 
    differences for each laser. Only when the difference is becoming as 
    small as a few Hz, the correlation would show up in the data. This only 
    happens for a few hours in total for all lasers per configuration for 
    the whole duration of more than 10 years and causes negligible extra  
    phase noise. The frequency plan considered here is computed for Earth-trailing orbits provided by \ac{ESA}.}
\end{figure}

Third, we consider possible correlations between the 2 \acp{ISI} and the 4 remaining interferometers on one \ac{SC}.
To minimize crosstalk in 
the 
\ac{DPLL}, the frequency plan ensures that the \acp{RFI}/\acp{TMI} and the \acp{ISI} on one \ac{SC} do not share the same heterodyne frequency, with a 
margin of about $\SI{2}{MHz}$. Therefore, no direct correlations are expected to occur. The 
remaining possibility is for correlations between 1f- and 2f-RIN, e.g., if one beat is at 
$\SI{12}{MHz}$ and the other one at $\SI{6}{MHz}$. This would lead to a 
correlation between 1f-RIN from the $\SI{12}{MHz}$ beat with 2f-RIN in the 
$\SI{6}{MHz}$ beat. However, this can be considered unproblematic, here again, as the 
1f-RIN term is likely to be dominating. In addition, such a scenario is a rare event: even with 
a 
relatively large threshold of $\SI{50}{Hz}$ difference between the beats 
(compared to the $\sim\SI{4}{Hz}$ measurement band), the maximal duration 
involving correlated measurements across all lasers and interferometers is in the order of a few hours for over 10 years of 
simulated frequency plan data (shown in \cref{fig:locking_correlations} 
for the baseline N1-L12 
configuration). Furthermore, the frequency plan could be further optimized to 
avoid such crossings, if desired.

%
%
%

 \subsection{Influence of \ac{TDI}}
 \ac{TDI} strongly suppresses laser phase noise by about 8 orders of 
 magnitude. For this 
 purpose, 
 it uses time shifted 
 combinations of the 
 interferometric phase measurements. As such, it also suppresses the laser locking noise ``echos'', since they appear in the measurements identically to the 
 laser frequency 
 noise that \ac{TDI} is designed to suppress.
 Unfortunately, this process also adds other noises from the 18 interferometric measurements into the resulting \ac{TDI} variables. 
 These noises have been studied and are now well understood (see, for example, \cite{tdi_noiseTF2022}). To first order, \ac{RIN} can be considered to behave like any other 
 uncorrelated readout noise due to its properties discussed before; 
 especially since the most significant \ac{RIN} contribution appears 
 uncorrelated in the \acp{ISI}, while the correlated appearances (\ac{TMI}, 
 \ac{RFI}) produce  
 much smaller noise contributions to the total measurement chain. According 
 to 
 \cite{tdi_noiseTF2022}, an uncorrelated
 readout noise (e.g., in units of 
 $\SI{}{m/\rtHz}$) entering all \acp{ISI} with a level of $\tilde{\varphi}$, 
 which describes the dominant \ac{RIN} contribution, has a \ac{PSD} in the 
 \ac{TDI} combination $X_2$ of
 \begin{align}\label{eqn:tdi_uncorr_isi}
     S_{X_{2}}(\omega, \tilde{\varphi}) &= 4 \tilde{\varphi}^2 
     C_{XX}(\omega) ,\\
 C_{XX}(\omega) &= 16 \sin[2](\omega\frac{L}{c}) \sin[2](\omega\frac{2 
 L}{c}),
 \end{align}
where $\omega = 2 \pi f$, and $c$ is the speed of light in a vacuum.

\section{Simulation architecture}
To verify the validity of the analytical derivations presented in the 
previous section and track the effects of possible correlations, we 
implemented a time-domain simulation.

\begin{figure}
    \includegraphics[width=0.5\textwidth]{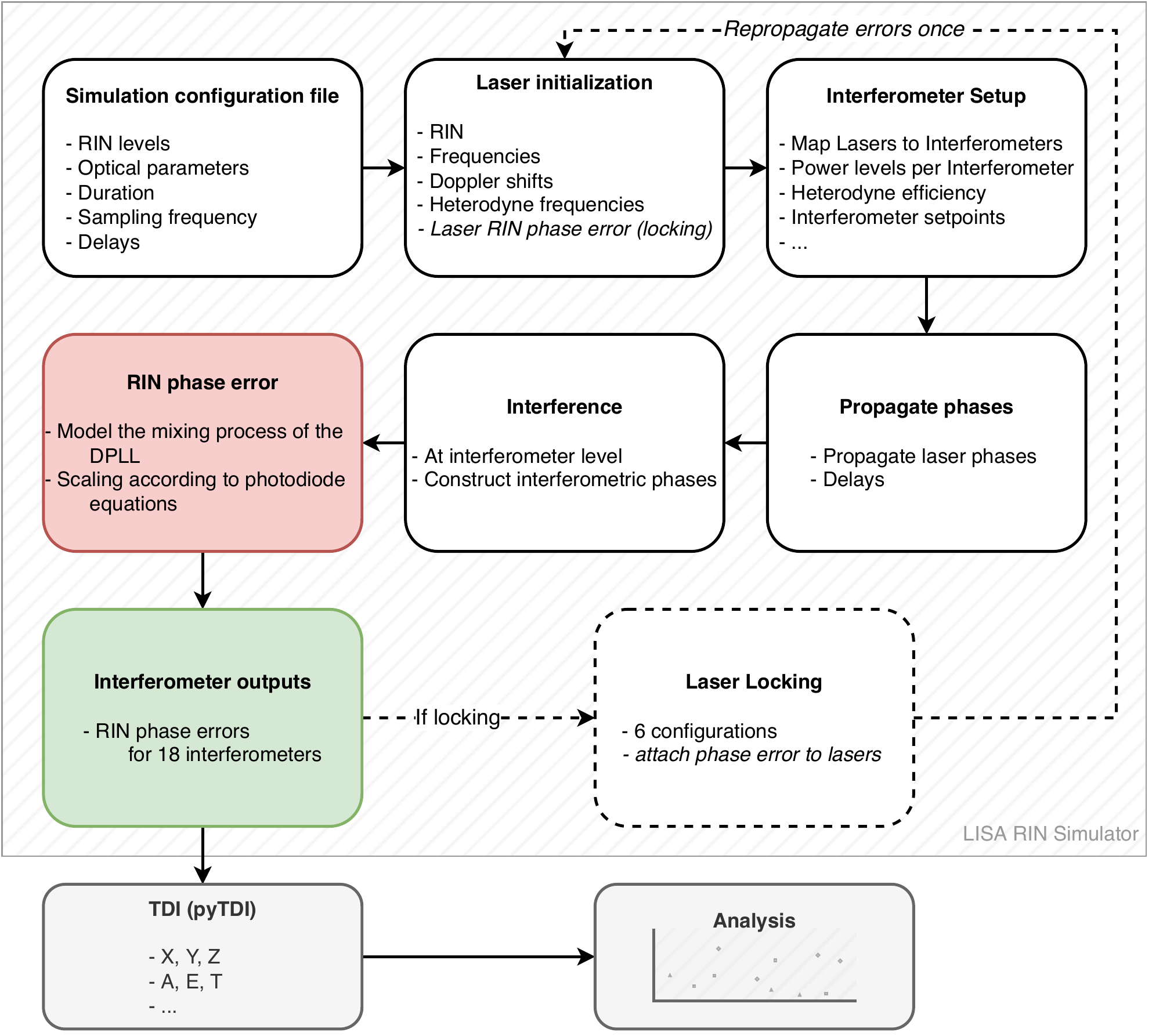}
    \caption{\label{fig:simulation_scheme} Diagram to show the different 
        stages of the \ac{RIN} simulation.}
\end{figure}

\Cref{fig:simulation_scheme} gives an overview of this 
simulator, which has multiple stages and is able to simulate the whole 
\ac{LISA}  
constellation with its 18 interferometers.
The simulation is performed in phase domain to easily represent the mixing 
process inside the \ac{DPLL}. The laser and optical parameters are read from 
a configuration file and then propagated to the interferometers, where the 
interference phase of two beams is simulated. Then, the \ac{RIN} phase 
error due to the mixing and demodulation process is calculated at high sampling 
frequency, and used as an output either for the locking scheme or directly 
for the output of the interferometers.
Consecutive scripts perform the data analysis tasks.

Typically, we simulate tens of thousands of seconds with a 
sampling frequency of $\SI{1}{kHz}$.

We simulate the total phase of each 
interferometer. Since we cannot simulate MHz frequencies directly for long  
time periods, we choose the offset frequencies such that all beatnote 
frequencies are in a range
between $\SI{100}{Hz}$ and $\SI{250}{Hz}$, instead of $\SI{5}{\mega\hertz}$ 
and $\SI{25}{\mega\hertz}$. This is sufficient to accurately model 
the \ac{RIN} coupling in the time domain, as its behavior is 
independent of the absolute heterodyne frequency.

We simulate laser locking, and correctly keep track of the beatnote 
polarities. The locking control loop is assumed to be perfect, such that the 
locking interferometer error signal is sign-inverted and added to the locked laser, 
and then propagates to all interferometers involving this laser. 

Following previous considerations for the frequency planning, we assume that 
only local interferometers on the same \ac{SC} share the same heterodyne 
frequency. Additionally, each laser carries its own \ac{RIN} noise 
time series, which is 
propagated (and delayed where applicable) to the corresponding 
interferometers.

The delays,  in the order of 
$\SI{8}{s}$, are constant 
and symmetric for the two directions of each arm, but not equal between 
different arms. The initial assumption of unequal beatnote frequencies would 
not be 
violated for slowly-varying arm lengths due to the frequency locking, as 
explained above. This is especially valid for our relatively short simulations, for which the frequencies do not change much.

The \ac{RIN} mixing is then applied using all beatnote phases and 
corresponding lasers with their \ac{RIN} time series. We include a model for 
balanced detection with 
different balancing efficiencies. The \ac{RIN} phase error is propagated as 
its own 
time series through the constellation, to avoid numerical problems with the  
large phase ramps of the beatnote phases. The outputs are filtered and 
downsampled, typically to a final output sampling frequency of 
$\SI{10}{Hz}$.

\section{Results}

\begin{figure*}
    \includegraphics[width=1\textwidth]{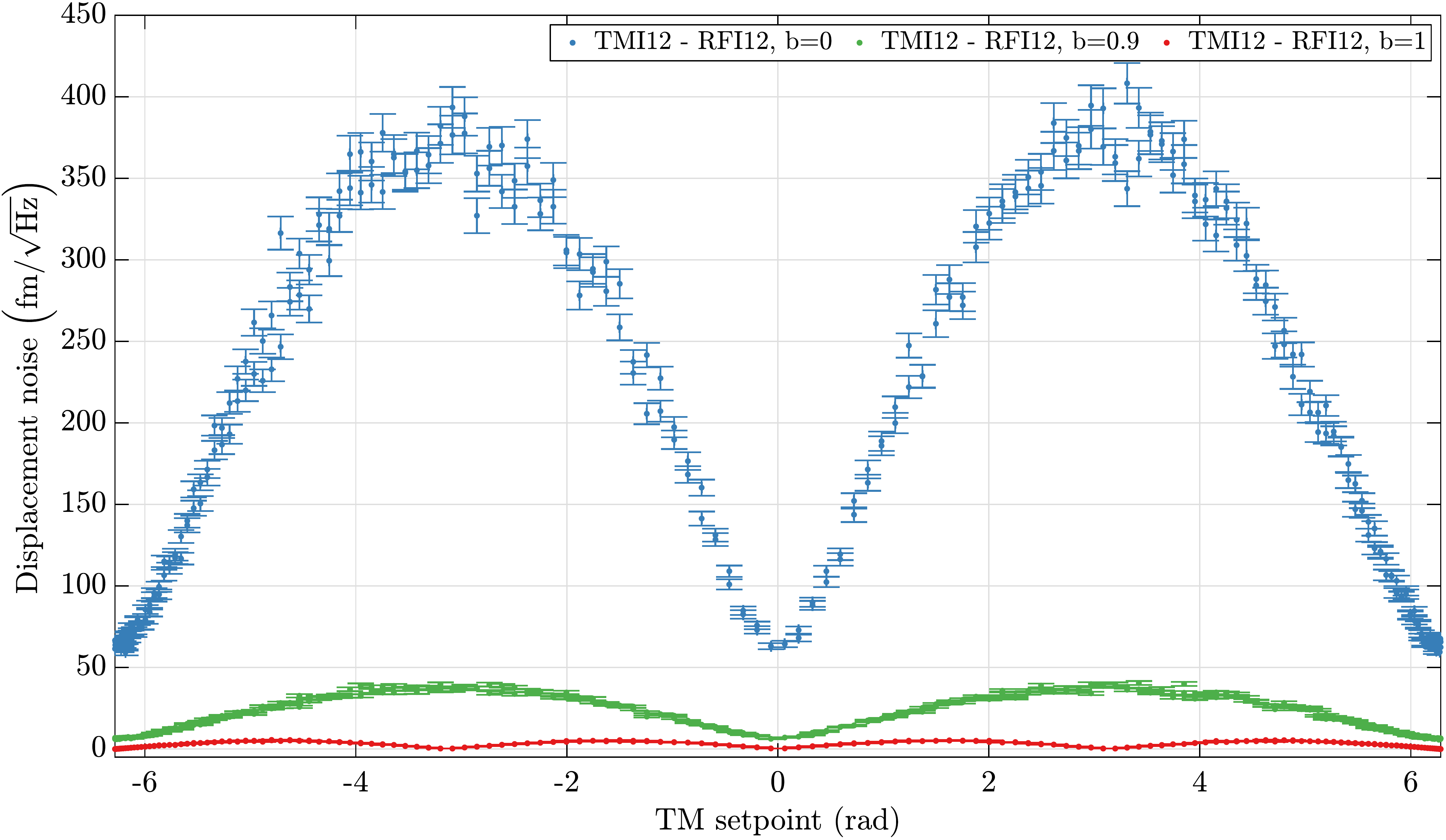}
    \caption{\label{fig:tmi_rfi} Simulated data for one local 
        $\text{TMI}_{ij} - \text{RFI}_{ij}$ subtraction with different balancing 
        efficiencies, in the unlocked case. This shows the 
        possible correlation properties of the 
        \ac{RIN} phase error, following the expected sinusoidal shape. For 
        perfect 1f-RIN subtraction $(b=1)$, only 2f-RIN 
        remains. The different noise floors are due to the unmatched power 
        levels in the two interferometers. For these simulations, a slow sine 
        injection from 
        $[-2\pi,2\pi]$ has been injected to mimic \ac{TM} motion for 
        $\SI{30000}{s}$ of 
        data sampled at 
        $\SI{1}{kHz}$. Each point shown here corresponds to the total 
        measurement having been cut into 
        $\SI{100}{s}$ segments, and their flat \acp{PSD} averaged between 
        \num{1} and \SI{3}{\hertz}.}
\end{figure*}

\begin{figure*}
    \includegraphics[width=1\textwidth]{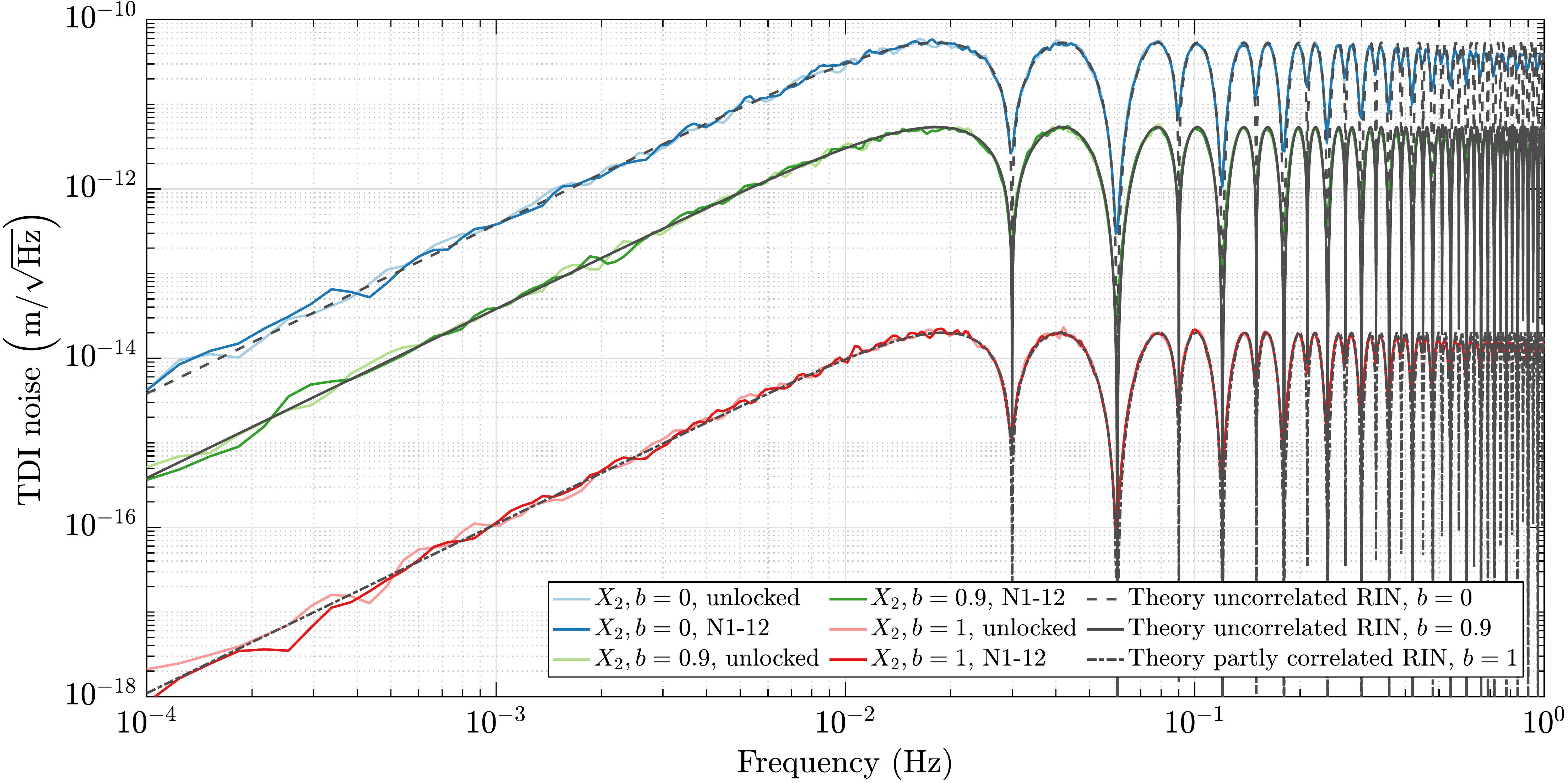}
    \caption{\label{fig:tdi_results} Propagation of the \ac{RIN} phase error 
        through \ac{TDI}, with a simulation duration of $\SI{1e5}{s}$ and $f_s = 
        \SI{1000}{Hz}$, for different balancing efficiencies and a comparison 
        between unlocked lasers and the N1-12 locking scheme. We see that the 
        \ac{RIN} phase error behaves like an \ac{ISI} uncorrelated readout 
        noise. This 
        plot has been produced with the lpsd algorithm 
        \cite{lpsd2006}. The theoretical expectations are plotted without the 
        usual relaxation towards lower frequencies.}
\end{figure*}

In this section, we present our analytical and simulated results for 
the \ac{LISA} mission parameters. The main findings are summarized in 
\cref{t:numerical_estimates}, \cref{fig:tmi_rfi} and \cref{fig:tdi_results}. We show the expected noise levels per interferometer, local common-mode 
suppression, and the propagation through \ac{TDI}.

\begin{table*}
    \begin{center}
        \begin{tabularx}{1\linewidth}{llccccc}
            \toprule 
            \textbf{Locking config.} &
            \textbf{Interferometer} & \textbf{1f-RIN} & 
            \textbf{2f-RIN} & \textbf{Total $(b=0.9)$} & \textbf{Total 
                $(b=1)$} & 
            \textbf{Total $(b=0)$} \\ 
            \midrule
            None & $\text{ISI}_{ij}$ & $\SI{8.7e-12}{}$ & $\SI{2.5e-15}{}$ & 
            $\SI{0.87e-12}{}$ & $\SI{2.5e-15}{}$ & $\SI{8.7e-12}{}$\\
            &$\text{TMI}_{ij}$ & $\SI{0.16e-12}{}$ & $\SI{2.5e-15}{}$ &
            $\SI{16.2e-15}{}$& $\SI{2.5e-15}{}$ & $\SI{0.16e-12}{}$\\
            &$\text{RFI}_{ij}$ & $\SI{0.22e-12}{}$ & $\SI{2.5e-15}{}$ &
            $\SI{22.2e-15}{}$& $\SI{2.5e-15}{}$ & $\SI{0.22e-12}{}$\\
            &$\text{TMI}_{ij} - \text{RFI}_{ij}$ &  $\SI{65.6e-15}{}$ & 
            $\SI{3.0e-16}{}$
            & $\SI{6.5e-15}{}$& $\SI{3.0e-16}{}$ & 
            $\SI{65.6e-15}{}$\\
            \midrule
            N1-12 & $\text{ISI}_{32}$ &  &  & 
            $\SI{1.5e-12}{}$ & $\SI{6.2e-15}{}$ & $\SI{15.0e-12}{}$\\
            & $\text{ISI}_{23}$ &  &  & 
            $\SI{1.5e-12}{}$ & $\SI{6.2e-15}{}$ & $\SI{15.0e-12}{}$\\
            & $\text{ISI}_{13}$ &  &  & 
            $\SI{1.2e-12}{}$ & $\SI{5.1e-15}{}$ & $\SI{12.3e-12}{}$\\
            \midrule
            N2-12 & $\text{ISI}_{23}$ &  &  & 
            $\SI{1.7e-12}{}$ & $\SI{6.2e-15}{}$ & $\SI{17.4e-12}{}$\\
            & $\text{RFI}_{32}$ &  &  & 
            $\SI{1.5e-12}{}$ & $\SI{6.2e-15}{}$ & $\SI{15.0e-12}{}$\\
            & $\text{RFI}_{31}$ &  &  & 
            $\SI{1.5e-12}{}$ & $\SI{6.2e-15}{}$ & $\SI{15.0e-12}{}$\\
            \midrule
            N3-12 & $\text{ISI}_{23}$ &  &  & 
            $\SI{1.7e-12}{}$ & $\SI{6.2e-15}{}$ & $\SI{17.4e-12}{}$\\
            & $\text{ISI}_{13}$ &  &  & 
            $\SI{1.5e-12}{}$ & $\SI{6.2e-15}{}$ & $\SI{15.0e-12}{}$\\
            & $\text{ISI}_{31}$ &  &  & 
            $\SI{1.5e-12}{}$ & $\SI{6.2e-15}{}$ & $\SI{15.0e-12}{}$\\
            \midrule
            N4-12 & $\text{ISI}_{32}$ &  &  & 
            $\SI{1.7e-12}{}$ & $\SI{7.2e-15}{}$ & $\SI{17.4e-12}{}$\\
            & $\text{RFI}_{21}$ &  &  & 
            $\SI{1.5e-12}{}$ & $\SI{6.2e-15}{}$ & $\SI{15.0e-12}{}$\\
            & $\text{RFI}_{23}$ &  &  & 
            $\SI{1.5e-12}{}$ & $\SI{6.2e-15}{}$ & $\SI{15.0e-12}{}$\\
            \midrule
            N5-12 & $\text{ISI}_{32}$ &  &  & 
            $\SI{1.7e-12}{}$ & $\SI{7.2e-15}{}$ & $\SI{17.4e-12}{}$\\
            & $\text{ISI}_{21}$ &  &  & 
            $\SI{1.5e-12}{}$ & $\SI{6.2e-15}{}$ & $\SI{15.0e-12}{}$\\
            & $\text{ISI}_{12}$ &  &  & 
            $\SI{1.5e-12}{}$ & $\SI{6.2e-15}{}$ & $\SI{15.0e-12}{}$\\
            \midrule
            N6-12 & $\text{ISI}_{31}$ &  &  & 
            $\SI{2.1e-12}{}$ & $\SI{8.0e-15}{}$ & $\SI{21.2e-12}{}$\\
            & $\text{ISI}_{23}$ &  &  & 
            $\SI{1.7e-12}{}$ & $\SI{6.2e-15}{}$ & $\SI{17.3e-12}{}$\\
            & $\text{RFI}_{13}$ &  &  & 
            $\SI{1.5e-12}{}$ & $\SI{6.2e-15}{}$ & $\SI{15.0e-12}{}$\\
            \bottomrule
        \end{tabularx}
    \end{center}
    \caption{Numerical simulation of the \ac{RIN}-to-phase 
        coupling in various interferometers, based on the 
        parameters listed in \cref{t:optical_params}. The first four rows show the 
        resulting phase noise in the unlocked case. 
        The following rows show the three interferometers with the largest 
        propagated phase noise for each locking configuration. The last 
        columns report 
        the total phase noise assuming 
        different balancing efficiencies (realistic case, best and worst). 
        The laser locking is assumed to be applied to the data after 
        balanced detection with the stated balancing efficiency.
        The numerical estimates are averages of the flat \ac{PSD} in the 
        band $\SIrange{0.01}{2}{Hz}$ (using 24 averages and 
        a Blackman-Harris92 window), from $\SI{30000}{s}$ of data sampled with 
        $\SI{1}{kHz}$ each.
        All 
        values are given in  
        $\si{\meter\hertz\tothe{-0.5}}$. The relative standard deviation is of the order 
        of a few \%, values are rounded to the last digit.}
    \label{t:numerical_estimates}
\end{table*}
\begin{figure}
    \includegraphics[width=0.5\textwidth]{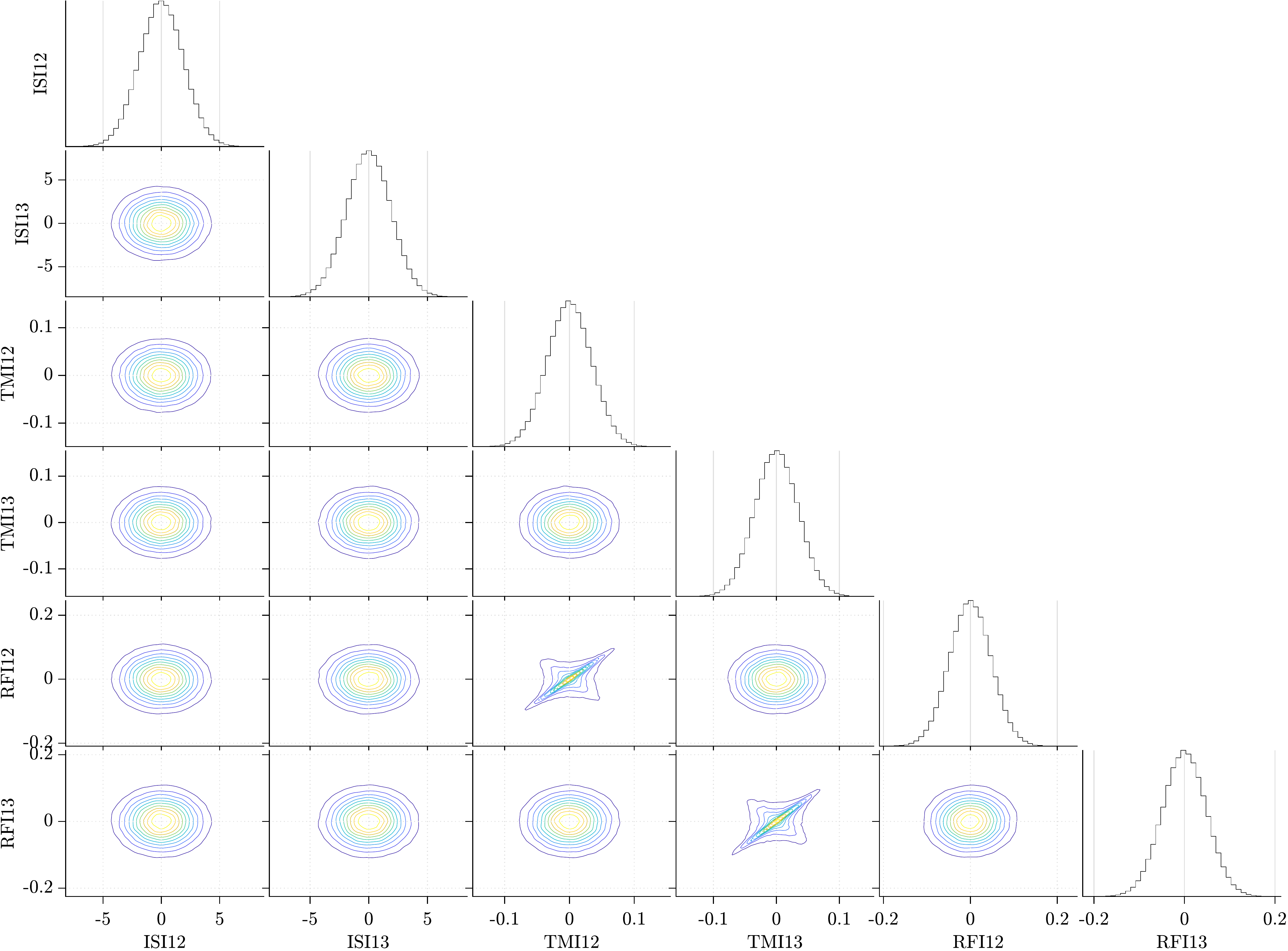}
    \caption{\label{fig:tmi_rfi_corr} Correlations between noises of the local 
        interferometers 
        from \cref{fig:tmi_rfi}, here for \ac{SC}~1, with $\SI{90}{\%}$ 
        balancing efficiency and in the unlocked case. The diagonal shows the histogram; off-diagonal elements 
        show density plots with units of picometers. The reason 
        for the slightly rectangular correlations between the local \ac{RFI} and 
        \ac{TMI} arises from 
        the mixing process, which contains a sinusoidal multiplication with a 
        phase modulation (the \ac{TM} setpoint) in this case. We see that the 
        \ac{RIN}-induced phase noise can be considered largely uncorrelated, 
        even between adjacent local interferometers, due to the inversely 
        distributed powers. The plot has been produced using \cite{corner}.}
\end{figure}

\subsection{1f-RIN estimates}
Based on the theoretical derivation above (and~\cite{wissel_rin2022}) and 
the optical parameters 
relevant for the \ac{RIN}-to-phase coupling from the current \ac{LISA} design (given in 
\cref{t:optical_params}), we estimate the expected noise 
levels in the 
three distinct interferometers without locking.

The 1f-RIN couplings are strong contributors to the phase noise in \ac{LISA}.
The \ac{TMI} shows a 1f-RIN contribution of about 
$\SI{155}{fm/\rtHz}$, the \ac{RFI} has a noise of about 
$\SI{220}{fm/\rtHz}$ and the \ac{ISI} reaches even a level of 
$\SI{8.7}{pm/\rtHz}$. Assuming a balancing efficiency of $\SI{90}{\%}$, 
these values reduce to $\SI{15.5}{fm/\rtHz},\, \SI{22}{fm/\rtHz}, \text{ and 
} 
\SI{0.87}{pm/\rtHz}$. 
The correlated $\text{TMI}-\text{RFI}$ subtraction in \ac{TDI} is able to 
reduce 
the contribution of these two interferometers to about $\SI{6.5}{fm/\rtHz}$. 
Contrary to the results presented in \cite{wissel_rin2022}, a complete 
subtraction is not possible in \ac{LISA} due to the 
unequal beam powers in the correlated interferometers, even if the residual 
translational \ac{SC} jitter $dx_\text{SC}$ vanishes.

In total, we get for a single 
link 
(uncorrelated \ac{ISI}, two uncorrelated $\text{TMI}-\text{RFI}$ 
measurements) $\sqrt{
    (\SI{0.87}{pm/\rtHz})^2 + (\sqrt{2}\cdot\SI{6.5}{fm/\rtHz})^2 }$, that is approximately $  
    \SI{0.87}{pm/\rtHz}$.

\subsection{2f-RIN estimates}
We find white-noise baseline estimates for 2f-RIN of about 
$\SI{2.5}{fm/\rtHz}$ in the \ac{ISI}, \ac{TMI} and \ac{RFI}. 
This value is expected to be identical across the interferometers, because 
the coupling is independent 
of beam parameters, such as powers.
Since it is correlated in the $\text{TMI}-\text{RFI}$ subtraction performed 
in \ac{TDI}, the contribution of these two interferometers is further 
reduced by a similar sine factor (with twice its argument) as the 1f-RIN 
 only adds marginal noise to the single link \ac{TM}-to-\ac{TM} 
measurement~\cite{wissel_rin2022}. Here, full subtraction is possible, since 
the noise in the 
correlated interferometers does not depend on the beam powers.

The total 2f-RIN noise in a single \ac{TM}-to-\ac{TM} link can be estimated  
by $\sqrt{(\SI{2.5}{fm/\rtHz})^2 + (\sqrt{2}\cdot 
    \SI{0.3}{fm/\rtHz})^2} \approx \SI{2.6}{fm/\rtHz}$.
Therefore, the phase noise due to 2f-RIN is much weaker than the phase noise 
caused by 1f-RIN.

\subsection{\ac{DWS}}
\ac{DWS} uses pairs of photodiode quadrants to sense wavefront tilts 
between the measurement and reference beams. It is used for \ac{SC} and 
\ac{TM} angular 
control, since the wavefront tilts can be calibrated to yield physical 
\ac{TM}-to-\ac{SC} angles. As such, it uses the same phase readout (yet different 
quadrant combinations, see for example \cite{heinzel_newDWS2020}) as the 
longitudinal channels and will therefore also be affected by \ac{RIN}. The 
behavior is expected to be the same as in the longitudinal 
$\text{TMI}-\text{RFI}$ common mode suppression (because pairs of 
quadrants are always combined), but with better results 
due to effectively equal power levels across the quadrants.
Since the expected angles measured through \ac{DWS} are rather small 
(usually less than $\SI{1}{rad}$ for the non-calibrated quadrant phase) and the 
quadrants 
share the same correlated 
\ac{RIN}, common-mode suppression, together with balanced detection, are 
expected to yield good minimization effects.

\subsection{Contribution summary}
In total, we find the quadratic sum of the 1f-RIN and 2f-RIN 
contributions for a single \ac{TM}-to-\ac{TM} link, with $\SI{90}{\%}$ 
balancing efficiency 
and no laser locking, to be at a level 
of $\SI{0.87}{pm/\rtHz}$ for the longitudinal readout. This amount of noise 
has to be considered as an 
entry in the $\sim \SI{10}{pm/\rtHz}$ noise budget of the total optical 
metrology noise. Let us note how important it is that the 
balancing requirements are met; if not, the total 
optical metrology budget would already be dominated by the \ac{RIN}-induced phase noise.

More detailed estimates are given in 
\cref{t:numerical_estimates} as a summary of the expected noise \ac{ASD} 
levels per local interferometer and the biggest local noise 
measurements for all locking schemes, based on the parameters of 
\cref{t:optical_params}. These have been calculated both analytically (where 
directly possible) and 
numerically simulated using the \ac{RIN} simulator described above, 
where we disabled all other noises.

\subsection{Local common-mode rejection}
In \cref{fig:tmi_rfi}, we present the local correlated behavior of 
the frequency-averaged \ac{RIN} phase error in the $\text{TMI}_{ij} - 
\text{RFI}_{ij}$ 
subtraction. The results follow the expected theoretical pattern (also 
agrees with \cref{t:numerical_estimates}) and show the common-mode rejection 
for a differential phase (labeled \ac{TM} setpoint here) between the two 
interferometers. The theoretical pattern is described in 
\cite{wissel_rin2022}. 
An important observation is that even for the correlated subtraction on the 
same \ac{MOSA}, the noise cannot be fully reduced due to the unequal beam 
powers. Note that this rejection would be even weaker if two 
interferometers from adjacent optical benches would be used in the 
subtraction, due to the even more unequal beam powers.
\Cref{fig:tmi_rfi_corr} shows the measured correlations in this simulation.

Similar effects have been observed in \ac{LPF} and on ground 
\cite{wissel_rin2022}. During the 
mission, a set-point close to $\SI{0}{rad}$ could be chosen to further 
minimize 
the noise.

\subsection{Propagation through \ac{TDI}}
Having simulated all interferometers with and without locking, the phase 
outputs need to be propagated through \ac{TDI}, similar to the real mission 
data. For this purpose we are using the software package PyTDI 
\cite{staab_martin_2022_6867012}.

In \cref{fig:tdi_results}, we show the results with different balancing 
efficiencies for the baseline locking configuration (N1-12), and compare these scenarios to 
the case of  
unlocked lasers.
We also overlay the analytical expectations given 
below. 
As expected, we find that \ac{TDI} suppresses the additional noises due to laser locking, and 
that the final noise resembles an uncorrelated readout 
noise, with the transfer function given in 
\cref{eqn:tdi_uncorr_isi}. The relevant noise level is given by
\begin{equation}
    \tilde{\varphi} = \tilde{\varphi}_\text{ISI} \approx (1-b) \cdot 
    \SI{8.7}{\frac{pm}{\rtHz}},
\end{equation}
with the condition that the \acp{ISI} dominate, and 
$b<1$. The residual 2f-RIN terms of $\SI{2.5}{fm/\rtHz}$ only have to be 
considered 
for perfect balanced detection ($b=1$, see below), when the correlated noise 
in the \ac{TMI} and \ac{RFI} also becomes relevant.

A more detailed (yet still simplified) upper bound for the total noise after 
\ac{TDI} can be estimated by adding 
the transfer function of the individual interferometer contributions, 
assuming the \ac{TMI} and \ac{RFI} add uncorrelated noise only. This 
leads to $\sqrt{2}$ smaller noise than the maximal possible contribution due 
to their correlation, but assuming good suppression due to the 
$\sin(dx_\text{SC})$ factor, this would still be a reasonably high upper 
bound that reduces the required estimation effort drastically.  
The total \ac{RIN} propagation after \ac{TDI} would then follow (using the 
results from \cite{tdi_noiseTF2022}),
\begin{align}
    S_{X_{2}} &= S_{X_{2},\text{ISI}} + 
    S_{X_{2},\text{TMI}} + S_{X_{2},\text{RFI}},
    \intertext{where}
    S_{X_{2},\text{ISI}} &= 4 C_{XX}(\omega) \cdot 
    \tilde{\varphi}_\text{ISI}^2, \\
    S_{X_{2},\text{RFI}} &= 4 C_{XX}(\omega) \cdot 
    \tilde{\varphi}_\text{RFI}^2, \\
    S_{X_{2},\text{TMI}} &= C_{XX} (\omega)
    \left(3+\cos(\omega \frac{2L}{c})\right) \cdot 
    \tilde{\varphi}_\text{TMI}^2 .
    \intertext{The noise levels according to the results from the 
    derivation in the previous sections,}
   \tilde{\varphi}_\text{IFO} &= \frac{\lambda}{2\pi}\sqrt{\left((1-b)  \cdot
   a_\text{IFO,1f} \cdot \tilde{n}_\text{1f}\right)^2 + \left(\frac{1}{2} 
   \tilde{n}_\text{2f}\right)^2}.
\intertext{This translates for the three distinct interferometers (assuming 
50/50 beamsplitters) to}
a_\text{ISI,1f} &= 
\frac{1}{\eta_\text{carrier}}\sqrt{\frac{P_{\text{ISI},1}^2 
        + 
        P_{\text{ISI},2}^2}{2 
        \eta_{\text{het,ISI}}  P_{\text{ISI},1} 
        P_{\text{ISI},2}}},  \\
   a_\text{RFI,1f} &= 
   \frac{1}{\eta_\text{carrier}}\sqrt{\frac{P_{\text{RFI},1}^2 
           +            P_{\text{RFI},2}^2}{2 
           \eta_{\text{het,RFI}}  P_{\text{RFI},1} 
           P_{\text{RFI},2}}},  \\
   a_\text{TMI,1f} &= 
   \frac{1}{\eta_\text{carrier}}\sqrt{\frac{P_{\text{TMI},1}^2 
           +            P_{\text{TMI},2}^2}{2 
           \eta_{\text{het,TMI}}  P_{\text{TMI},1} 
           P_{\text{TMI},2}}},
   \intertext{such that we find for the noise levels}
   \tilde{\varphi}_\text{ISI} &\approx (1-b) \cdot 
   \SI{8.7}{\frac{pm}{\rtHz}}, \\
   \tilde{\varphi}_\text{RFI} &\approx (1-b) \cdot 
   \SI{220}{\frac{fm}{\rtHz}}, \\
   \tilde{\varphi}_\text{TMI} &\approx (1-b) \cdot 
   \SI{155}{\frac{fm}{\rtHz}},
\end{align}
which is clearly dominated by the \ac{ISI} terms. 

However, in the limit of $b \rightarrow 1$, the 2f-RIN contributions and 
their correlations
 become relevant. Then, the \ac{RIN} residual is expected to be dominated by 
 the sum of uncorrelated contributions in the \acp{ISI} and the fully 
 correlated contribution among the local interferometers (\ac{TMI} and 
 \ac{RFI}) in the two adjacent \acp{MOSA}. This special case of 2f-RIN  
 correlation is not discussed in the literature, hence we give the 
 derivation here.

If we assign the same noise to all local interferometers on one spacecraft 
($\text{RFI}_{ij} = \text{RFI}_{ik} = \text{TMI}_{ij} = \text{TMI}_{ik} = 
n_{i,\text{2f-RIN}} $) and perform a derivation similar to that presented in 
\cite{tdi_noiseTF2022} (assuming equal arm lengths), we recover the 
following residual in 
$X_{2}$,
\begin{equation}
    X_{2,\text{2f-RIN,corr.}} = (1 - D^2)^2(1 - D^4) 
    \varphi_{1,\text{2f-RIN}}.
\end{equation}
We note 
that only the \ac{RIN} contribution of \ac{SC}1 remains in $X_{2}$, while 
those of the other two \ac{SC} cancel. We compute the
\ac{PSD} by taking the Fourier transform of the previous equation and 
calculating the expectation value of the squared magnitude, yielding
\begin{equation}
    S_{X_{2},\text{2f-RIN,corr.}} = 4 \sin[2](\omega \frac{L}{c}) 
    C_{XX}(\omega) \cdot 
    \tilde{\varphi}_\text{2f-RIN}^2,
\end{equation}
where $\tilde{n}_\text{2f-RIN}$ is the equal level of 2f-RIN in all 
interferometers and the usual \ac{TDI} transfer function is modulated by an 
additional sine squared factor. This causes a faster roll-off of the  
\ac{PSD} towards low frequencies and is thus only relevant at the maxima of 
the 
\ac{TDI} transfer function.
The sum of the uncorrelated \ac{ISI} \ac{RIN} $S_{X_{2},\text{ISI}}$ term 
for $b=1$ and the locally
correlated  2f-RIN $S_{X_{2},\text{2f-RIN,corr.}}$ term is plotted 
in \cref{fig:tdi_results} and agrees well with the simulation result.

In a nutshell, the 
simulated noise coupling propagation through \ac{TDI} are in very good 
agreement with 
the theoretical predictions, and the correlations do not have any 
significant influence under realistic circumstances.


\section{Conclusions}
We have analyzed, derived, and simulated the \ac{RIN}-to-phase noise coupling 
in \ac{LISA}, a future gravitational-wave observatory in space, and present 
for the first time a complete analysis of the influence of \ac{RIN} in the  
interferometric readout. We have considered the mission characteristics such as laser properties, 
optical bench design, and orbital dynamic influences, as well as mitigation 
strategies.

We conclude that the resulting phase noise follows our theoretical 
understanding and experience from \acf{LPF}. It is well under control for 
the 
current design parameters of reasonably low input \ac{RIN} and strong 
suppression of the dominating 1f-RIN through sufficient balanced detection.

With the mitigation strategies considered here, the \ac{RIN} phase noise is 
of the order of $\SI{0.87}{pm/\rtHz}$ for the single link \ac{TM}-to-\ac{TM} measurement along one \ac{LISA} arm, and dominated by the 1f-RIN 
terms in the inter-\ac{SC} interferometers. The 2f-RIN coupling only plays a 
subdominant role.

This amount of noise is below other secondary noises that are in 
the order up to a few \si{\pico\meter/\rtHz}.

Due to Doppler shifts and frequency planning, the \ac{RIN} coupling can 
essentially be considered as an uncorrelated readout noise, and behaves as 
such when it is propagated through \ac{TDI}.

We confirm that an additional mitigation of 1f- and 2f-RIN in the local 
$\text{TMI}_{ij} - 
\text{RFI}_{ij}$ can be achieved by choosing an interferometric operating 
point close to $\SI{0}{rad}$. This, however, cannot lead to perfect 
cancellation due to the unequal power levels.

In the \acp{DWS}, both 1f-RIN and 2f-RIN are expected to be strongly suppressed due to the recombination of correlated 
neighboring quadrants with almost equal powers, and therefore good 
common-mode suppression characteristics.

Future work may be focused on verifying our results using actual hardware 
representative of the real \ac{LISA} system, as 
well as failure studies and the more detailed analysis of the physical 
effects influencing the balancing efficiency.

\begin{acknowledgements}
   The authors would like to thank M. Misfeldt for his insightful comment.
   
   The Albert Einstein Institute gratefully acknowledges the support of the 
   German Space Agency, DLR. 
   The work is supported by the Federal Ministry for Economic Affairs and 
   Energy based on a resolution of the German Bundestag (FKZ 50OQ0501, FKZ 
   50OQ1601 and FKZ 50OQ1801).
   
   J.B.B. gratefully acknowledges support from UK Space Agency (grant 
   ST/X002136/1).
   UKATC also acknowledges support from the UK Space Agency.
   
   O.H. gratefully acknowledges support by Centre National d'\'Etudes Spatiales (CNES) and by the Programme National GRAM of CNRS/INSU with INP and IN2P3 co-funded by CNES.
\end{acknowledgements}

\bibliography{bibs_sorted_2.bib}
\begin{acronym}[foo]

    \acro{ADC}{Analogue to Digital Convertor}
    \acro{AEI}{Albert Einstein Institute}
    \acro{ao}{Analysis Object}
    \acro{AOM}{Acousto-Optic Modulator}
    \acro{APC}{AstroParticule et Cosmologie, Universit\'e Paris Diderot}
    \acro{ASD}{Amplitude Spectral Density}
    \acro{ASU}{Astrium UK}
    \acro{BS}{Beamsplitter}
    \acro{BH92}{Blackmann-Harris-92}
    \acro{CAD}{Computer Aided Design}
    \acro{CLG}{Closed-Loop Gain}
    \acro{CLTF}{Closed-Loop Transfer Function}
    \acro{CMM}{Coordinate Measurement Machine}
    \acro{CQP}{Calibrated Quadrant Photodiode}
    \acro{DAC}{Digital to Analogue Convertor}
    \acro{DC}{Direct Current}
    \acro{DDS}{Data Disposition System}
    \acro{DFACS}{Drag-Free and Attitude Control System}
    \acro{DFT}{Discrete Fourier transform}
    \acro{DMU}{Data Management Unit}
    \acro{DPLL}{Digital Phase-Locked Loop}
    \acro{DOF}{Degree-of-Freedom}
    \acro{DOY}{Day of year}
    \acro{DPS}{Differential Power Sensing}
    \acro{DRS}{Disturbance Reduction System}
    \acro{DWS}{Differential Wavefront Sensing}
    \acro{eLISA}{Evolved LISA}
    \acro{ELITE}{European LIsa TEchnology}
    \acro{EM}{Engineering Model}
    \acro{EMP}{Experimental Master Plan}
    \acrodefplural{EMs}{Engineering Models}
    \acro{ESA}{European Space Agency}
    \acro{FF}{Fast Frequency}
    \acro{FFT}{Fast Fourier transform}
    \acro{FIOS}{Fibre Injector Optical Sub-Assembly}
    \acro{FIR}{Finite Impulse Response}
    \acro{FM}{Flight Model}
    \acro{FP}{Fast Power}
    \acro{FPGA}{Field Programmable Gate Array}
    \acro{FT}{Fourier Transform}
    \acro{GRS}{Gravitational Reference Sensor}
    \acro{GUI}{Graphical User Interface}
    \acro{GW}{Gravitational Wave}
    \acro{IABG}{Industrieanlagen-Betriebsgesellschaft mbH}
    \acro{ICE}{Instrument Configuration Evaluation}
    \acro{IDL}{Interferometer Data Log}
    \acro{IFO}{Interferometer}
    \acro{IGR}{Institute for Gravitational Research, University of Glasgow}
    \acro{IIR}{Infinite Impulse Response}
    \acro{ISI}{Inter-Spacecraft Interferometer}
    \acro{ITO}{Indium Tin Oxide}
    \acro{KT}{Kaiser Threde}
    \acro{$L_1$}{Lagrange point 1}
    \acro{LA}{Laser Assembly}
    \acro{LA PFM}{Laser Assembly Pre-Flight Model}
    \acro{LCA}{LTP Core Assembly}
    \acro{LCU}{Laser Control Unit}
    \acro{LISA}{Laser Interferometer Space Antenna}
    \acro{LMU}{Laser Modulation Unit}
    \acro{LPF}{LISA Pathfinder}
    \acro{LPSD}{Logarithmic frequency axis Power Spectral Density}
    \acro{LTP}{LISA Technology Package}
    \acro{LTPDA}{LISA Technology Package Data Analysis}
    \acro{MAD}{Median Absolute Deviation}
    \acro{MBW}{measurement bandwidth}
    \acro{MCMC}{Markov Chain Monte-Carlo}
    \acro{MOC}{Mission Operations Centre}
    \acro{MOSA}{Moving Optical Sub-Assembly}
    \acro{NASA}{National Aeronautics and Space Administration}
    \acro{NCO}{Numerically Controlled Oscillator}
    \acro{NPRO}{Non-Planar Ring Oscillator}
    \acro{NTE}{NTE Sener with IEEC, Barcelona}
    \acro{OB}{Optical Bench}
    \acro{OBC}{Onboard Computer}
    \acro{OBF}{Optical Bench Frame}
    \acro{OBI}{Optical Bench Interferometer}
    \acro{OGSE}{On-ground Support Equipment}
    \acro{OLG}{Open-Loop Gain}
    \acro{OLTF}{Open-Loop Transfer Function}
    \acro{OMS}{Optical Metrology System}
    \acro{OPD}{Optical Pathlength Difference}
    \acro{OSTT}{On-Station Thermal Tests}
    \acro{OW}{Optical Window}
    \acro{PD}{Photodiode}
    \acro{PFM}{Pre-Flight Model}
    \acrodefplural{PFMs}{Pre-flight Models}
    \acro{PLL}{Phase Locked Loop}
    \acro{PM}{Phasemeter}
    \acro{PPS}{Pulse per second}
    \acro{PSD}{Power Spectral Density}
    \acro{PT}{Phase Tracking}
    \acro{PZT}{Piezo-electric Transducer}
    \acro{QPD}{Quadrant Photodiode}
    \acrodefplural{PZTs}{Piezo-electric Transducers}
    \acro{RAM}{Random Access Memory}
    \acro{RF}{Radio Frequency}
    \acro{RFI}{Reference Interferometer}
    \acro{RIN}{Relative Intensity Noise}
    \acro{RLU}{Reference Laser Unit}
    \acro{RMS}{Root Mean Square}
    \acro{RPN}{Relative Power Noise}
    \acro{SBDFT}{Single-Bin Discrete Fourier Transform}
    \acro{SC}{Spacecraft}
    \acro{SDM}{Science Data Mode}
    \acro{SDP}{System Data Pool}
    \acro{SEP}{Solar Energetic Particles}
    \acro{SEPD}{Single Element Photodiode}
    \acro{SF}{Slow Frequency}
    \acro{SF}{Slow Frequency}
    \acro{SID}{System Identification}
    \acro{SMART-2}{Small Missions for Advanced Research in Technology}
    \acro{SNR}{Signal to Noise Ratio}
    \acro{SP}{Slow power}
    \acro{SSC}{Source Sequence Counter}
    \acro{SSM}{State Space Model}
    \acro{SSMM}{Solid-state Mass Memory}
    \acro{ST-7}{Space Technology-7}
    \acro{STOC}{Science Technology Operations Centre} 
    \acro{SVN}{Small Vector Noise}
    \acro{TC}{Telecommand}
    \acro{TDI}{Time-Delay Interferometry}
    \acro{TIA}{Trans-Impedance Amplifier}
    \acro{TM}{Test Mass}
    \acrodefplural{TM}{Test Masses}
    \acro{TMI}{Test Mass Interferometer}
    \acro{TN}{Technical Note}
    \acro{TTL}{Tilt-To-Length}
    \acro{UOB}{University of Birmingham}
\end{acronym}

\end{document}